\documentclass{aa}

\usepackage{graphicx}
\usepackage{txfonts, longtable, dcolumn, ctable, lscape, multirow, lineno, bbding, pifont}
\usepackage[T1]{fontenc}
\usepackage{natbib}
\bibpunct{(}{)}{;}{a}{}{,}

\begin{document}

   \title{An anisotropic distribution of spin vectors in asteroid families}

   \author{J.~Hanu{\v s}
	  \inst{1*}
	   \and
          M.~Bro{\v z}\inst{1}
           \and
          J.~{\v D}urech\inst{1}
           \and
          B.~D.~Warner\inst{2}
	   \and
	  J.~Brinsfield\inst{3}
	  \and
	  R.~Durkee\inst{4}
	   \and
          D.~Higgins\inst{5}
           \and
          R.~A.~Koff\inst{6}
           \and
	  J.~Oey\inst{7}
	   \and
          F.~Pilcher\inst{8}
           \and
	  R.~Stephens\inst{9}
           \and
	  L.~P.~Strabla\inst{10}
           \and
	  Q.~Ulisse\inst{10}
           \and
	  R.~Girelli\inst{10}
   } 

   \institute{Astronomical Institute, Faculty of Mathematics and Physics, Charles University in Prague,
              V~Hole{\v s}ovi{\v c}k{\'a}ch 2, 18000 Prague, Czech Republic\\
              $^*$\email{hanus.home@gmail.com}
	 \and
             Palmer Divide Observatory, 17995 Bakers Farm Rd., Colorado Springs, CO 80908, USA 
         \and
	     Via Capote Observatory, Thousand Oaks, CA 91320, USA 
         \and
	     Shed of Science Observatory, 5213 Washburn Ave. S, Minneapolis, MN 55410, USA 
	 \and
	     Hunters Hill Observatory, 7 Mawalan Street, Ngunnawal ACT 2913, Australia 
	 \and
	     980 Antelope Drive West, Bennett, CO 80102, USA 
	 \and    
	     Kingsgrove, NSW, Australia 
	 \and
	     4438 Organ Mesa Loop, Las Cruces, NM 88011, USA 
	 \and
	     Center for Solar System Studies, 9302 Pittsburgh Ave, Suite 105, Rancho Cucamonga, CA 91730, USA
	 \and
	     Observatory of Bassano Bresciano, via San Michele 4, Bassano Bresciano (BS), Italy 
  }

   \date{Received x-x-2013 / Accepted x-x-2013}
 
  \abstract
   {Current amount of $\sim$500 asteroid models derived from the disk-integrated photometry by the lightcurve inversion method allows us to study not only the spin-vector properties of the whole population of MBAs, but also of several individual collisional families.}
   {We create a data set of 152 asteroids that were identified by the HCM method as members of ten collisional families, among them are 31 newly derived unique models and 24 new models with well-constrained pole-ecliptic latitudes of the spin axes. The remaining models are adopted from the DAMIT database or the literature.}
   {We revise the preliminary family membership identification by the HCM method according to several additional criteria -- taxonomic type, color, albedo, maximum Yarkovsky semi-major axis drift and the consistency with the size-frequency distribution of each family, and consequently we remove interlopers. We then present the spin-vector distributions for asteroidal families Flora, Koronis, Eos, Eunomia, Phocaea, Themis, Maria and Alauda. We use a combined orbital- and spin-evolution model to explain the observed spin-vector properties of objects among collisional families.}
   {In general, we observe for studied families similar trends in the ($a_\mathrm{p}$, $\beta$) space (proper semi-major axis vs. ecliptic latitude of the spin axis):
(i) larger asteroids are situated in the proximity of the center of the family;
(ii) asteroids with $\beta>0^{\circ}$ are usually found to the right from the family center;
(iii) on the other hand, asteroids with $\beta<0^{\circ}$ to the left from the center;
(iv) majority of asteroids have large pole-ecliptic latitudes ($|\beta|\gtrsim30^{\circ}$); and finally
(v) some families have a statistically significant excess of asteroids with $\beta>0^{\circ}$ or $\beta<0^{\circ}$.
Our numerical simulation of the long-term evolution of a collisional family is capable of reproducing well the observed spin-vector properties. Using this simulation, we also independently constrain the age of families Flora (1.0$\pm$0.5~Gyr) and Koronis (2.5--4~Gyr).}
   {}
 

  \titlerunning{Spin vectors in asteroid families}
  \maketitle

\section{Introduction}\label{introduction}

An analysis of rotational state solutions for main belt asteroids was performed by many authors. All the authors observed the deficiency of poles close to the ecliptic plane \citep[e.g.,][]{Magnusson1986,Drummond1988,Pravec2002,Skoglov2002,Kryszczynska2007}. \citet{Hanus2011} showed that this depopulation of spin vectors concerns mainly smaller asteroids ($D\lesssim40$ km), while the larger asteroids \citep[$60\lesssim D\lesssim$~130--150 km,][]{Kryszczynska2007, Paolicchi2012} have a statistically significant excess of prograde rotators, but no evident lack of poles close to the ecliptic plane. The observed anisotropy of pole vectors of smaller asteroids is now believed to be a result of YORP thermal torques\footnote{Yarkovsky--O'Keefe--Radzievskii--Paddack effect, a torque caused by the recoil force due to anisotropic thermal emission, which can alter both rotational periods and orientation of spin axes, see e.g., \citet{Rubincam2000}} and also collisions that systematically evolve the spin axes away from the ecliptic plane, and the prograde excess of larger asteroids as a primordial preference that is in agreement with the theoretical work of \citet{Johansen2010}. While the number of asteroids with known rotational states grows, we can study the spin vector distribution not only in the whole MBAs or NEAs populations, but we can also focus on individual groups of asteroids within these populations, particularly on collisional families (i.e., clusters of asteroids with similar proper orbital elements and often spectra that were formed by catastrophic break-ups of parent bodies or cratering events).

\begin{figure*}
\begin{center}
\resizebox{\hsize}{!}{\includegraphics{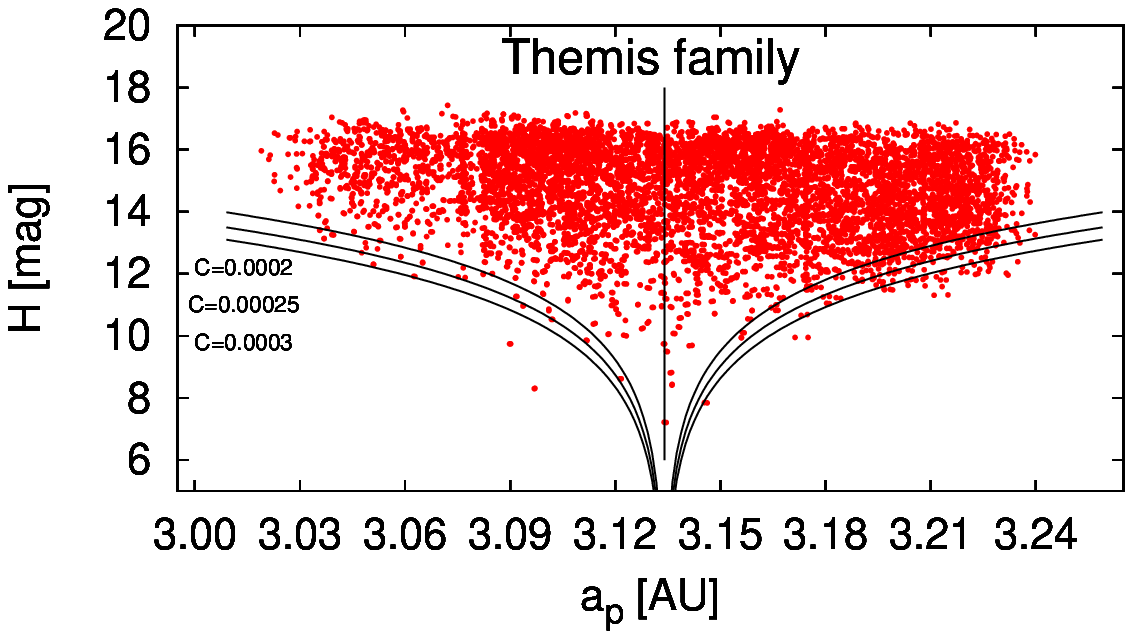}\,\includegraphics{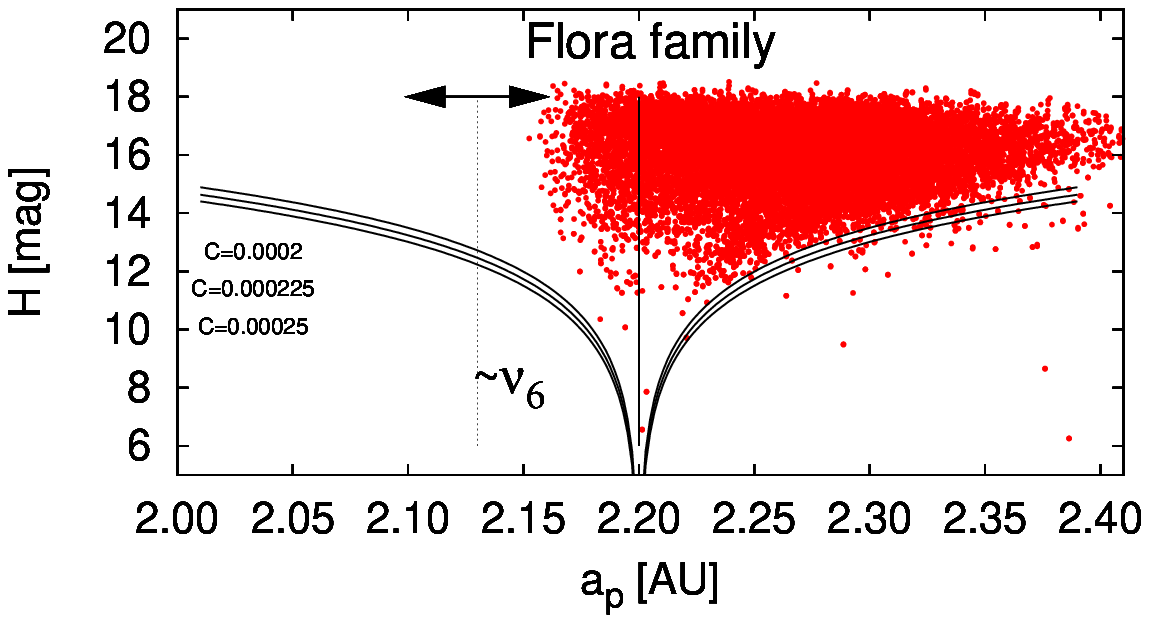}}\\
\end{center}
\caption{\label{img:a_H}Dependence of the absolute magnitude $H$ on the proper semi-major axis $a_{\mathrm{p}}$ for the Themis family (left) and for the Flora family (right) with the likely positions of the family centers (vertical lines). We also plot three ($a_{\mathrm{p}}$, $H$) borders of the family for different parameters $C$ (different values correspond to a different initial extent of the family or different age and magnitude of the Yarkovsky semi-major axis drift) by gray lines, the optimal border corresponds to the middle line. The vertical dotted line represents the approximate position of the secular $\nu_6$ resonance for the inclination typical for Flora family members and the horizontal arrow its approximate range.}
\end{figure*}

The theory of dynamical evolution of asteroid families \citep[e.g.,][]{Bottke2006} suggests that the Yarkovsky\footnote{a thermal recoil force affecting rotating asteroids}/YORP effects change orbital parameters of smaller asteroids ($\lesssim$30--50 km) -- the semi-major axis of prograde rotators is slowly growing in the course of time, contrary to retrograde rotators which semi-major axis is decreasing. This phenomenon is particularly visible when we plot the dependence of the absolute magnitude $H$ on the proper semi-major axis $a_{\mathrm{p}}$ (see an example of such plot for Themis family in Figure~\ref{img:a_H}, left panel). In addition, various resonances (e.g., mean-motion resonances with Jupiter or Mars, or secular resonances) can intersect the family and cause a decrease of the number of asteroids in the family by inducing moderate oscillations to their orbital elements\citep{Bottke2001} as can be seen in Figure~\ref{img:a_H} for the Flora family, where the secular $\nu_6$ resonance with Saturn almost completely eliminated objects to the left from the center of the family (the $\nu_6$ resonance has its center at 2.13 AU for objects with $\sin I\sim0.09$, which is typical for Flora family members, it evolves objects that come to the proximity of the resonance). Some resonances can, for example, capture some asteroids on particular semi-major axes \citep{Nesvorny1998}.

Laboratory experiments strongly suggest that a collisionally-born cluster should initially have a rotational frequency distribution close to Maxwellian \citep{Giblin1998} and an isotropic spin vector distribution.

For several families, we already know their age estimates \citep[e.g., $2.5\pm1.0$~Gyr for Koronis family, ][]{Bottke2001}, and so we have a constraint on the time, for which the family was evolving towards its current state. As was shown in \citet{Bottke2001}, the family evolution is dominated by Yarkovsky and YORP effects, and also collisions and spin-orbital resonances. The knowledge of the age should constrain some free parameters in various evolutionary models.

The spin-vector properties in an asteroid family were first studied by \citet{Slivan2002} and \citet{Slivan2003}, who revealed an anisotropy of spin vectors for ten members of the Koronis family. This was an unexpected result because collisionally-born population should have an isotropic spin-vector distribution. The peculiar spin-vector alignment in the Koronis family was explained by \citet{Vokrouhlicky2003} as a result of the YORP torques and spin-orbital resonances that modified the spin states over the time span of 2--3~Gyr. The secular $s_6$ spin-orbital resonance with Saturn may affect the Koronis family members, according to the numerical simulations, it can 
(i)~capture some objects and create a population of prograde rotators with periods $P\in(4,7)$~h, similar obliquities ($42^{\circ}$ to $51^{\circ}$) and also with similar ecliptic longitudes in the ranges of ($24^{\circ}$ to $73^{\circ}$) and ($204^{\circ}$ to $259^{\circ}$); or 
(ii)~create a group of low-obliquity retrograde rotators with rotational periods $P<5$~h or $P>13$~h. 
The prograde rotators trapped in the $s_6$ spin-orbital resonance were referred by \citet{Vokrouhlicky2003} as being in {\em Slivan states}. Most members of the Koronis family with known rotational states \citep[determined by the lightcurve inversion by][]{Slivan2003,Slivan2009,Hanus2011,Hanus2013a} had the expected properties except the periods of observed prograde rotators were shifted to higher values of 7--10~h. Rotational states of asteroids that did not match the properties of the two groups were probably reorientated by recent collisions, which are statistically plausible during the family existence for at least a few Koronis members \citep[e.g., asteroid (832)~Karin was affected by a collision when a small and young collisional family within the Koronis family was born][]{Slivan2012}.

Another study of rotational states in an asteroid family was performed by \citet{Kryszczynska2013a}, who focused on the Flora family. She distinguished prograde and retrograde groups of asteroids and reported an excess of prograde rotators. This splitting into two groups is likely caused by the Yarkovsky effect, while the prograde excess by the secular $\nu_6$ resonance that significantly depopulates the retrograde part of the family (see Figure~\ref{img:a_H}b, only retrograde rotators can drift via the Yarkovsky/YORP effects towards the resonance).

Further studies of rotational properties of collisional families should reveal the influence of the Yarkovsky and YORP effects, and possibly a capture of asteroids in spin-orbital resonances similar to the case of the Koronis family. The Yarkovsky effect should be responsible for spreading the family in a semi-major axis (retrograde rotators drift from their original positions towards the Sun, on the other hand, prograde rotators drift away from the Sun, i.e. towards larger $a_{\mathrm{p}}$'s), and the YORP effect should eliminate the spin vectors close to the ecliptic plane. 

Disk-integrated photometric observations of asteroids contain information about object's physical parameters, such as the shape, the sidereal rotational period and the orientation of the spin axis. Photometry acquired at different viewing geometries and apparitions can be used in many cases in a lightcurve inversion method \citep[e.g.,][]{Kaasalainen2001a,Kaasalainen2001b} and a convex 3D shape model including its rotational state can be derived. This inverse method uses all available photometric data, both the classical dense-in-time lightcurves or the sparse-in-time data from astrometric surveys. Most of the asteroid models derived by this technique are publicly available in the Database of Asteroid Models from Inversion Techniques \citep[DAMIT\footnote{\texttt{http://astro.troja.mff.cuni.cz/projects/asteroids3D}},][]{Durech2010}. In February 2013, models of 347 asteroids were included there. About a third of them can be identified as members of various asteroid families. This high number of models of asteroids that belong to asteroid families allows us to investigate the spin-vector properties in at least several families with the largest amount of identified members. Comparison between the observed and synthetic (according to a combined orbital- and spin-evolution model) spin-vector properties could even lead to independent family age estimates.

The paper is organized as follows: in Section~\ref{sec:models}, we investigate the family membership of all asteroids for which we have their models derived by the lightcurve inversion method and present 31 new asteroid models that belong to ten asteroid families. An analysis of spin states within these asteroid families with at least three identified members with known shape models is presented in Section~\ref{sec:spin_state_general}. A combined spin-orbital model for the long-term evolution of a collisional family is described in Section~\ref{sec:simulation}, where we also compare the synthetic and observed spin-vector properties and constrain ages of families Flora and Koronis.

\section{Determination of family members}\label{sec:models}

\subsection{Methods for family membership determination}\label{sec:membership}

For a {\em preliminary} family membership determination, we adopted an on-line catalog published by \citet{Nesvorny2012} who used the Hierarchical Clustering Method\footnote{In this method, mutual distances in proper semi-major axis ($a_{\mathrm{d}}$), proper eccentricity ($e_{\mathrm{d}}$), and proper inclination ($i_{\mathrm{d}}$) space are computed. The members of the family are then separated in the proper element space by less than a selected distance (usually, it has a unit of velocity), a free parameter often denoted as ``cutoff velocity``.} \citep[HCM,][]{Zappala1990,Zappala1994}. \citet{Nesvorny2012} used two different types of proper elements for the family membership identification: semi-analytic and synthetic. The more reliable dataset is the one derived from synthetic proper elements, which were computed numerically using a more complete dynamical model. Majority of asteroids is present in both datasets. A few asteroids that are only in one of the datasets are included in the study as well (e.g., asteroids (390)~Alma in the Eunomia family or (19848)~Yeungchuchiu in the Eos family), because at this stage it is not necessary to remove objects that still could be real family members.

\begin{figure}
\begin{center}
\resizebox{\hsize}{!}{\includegraphics{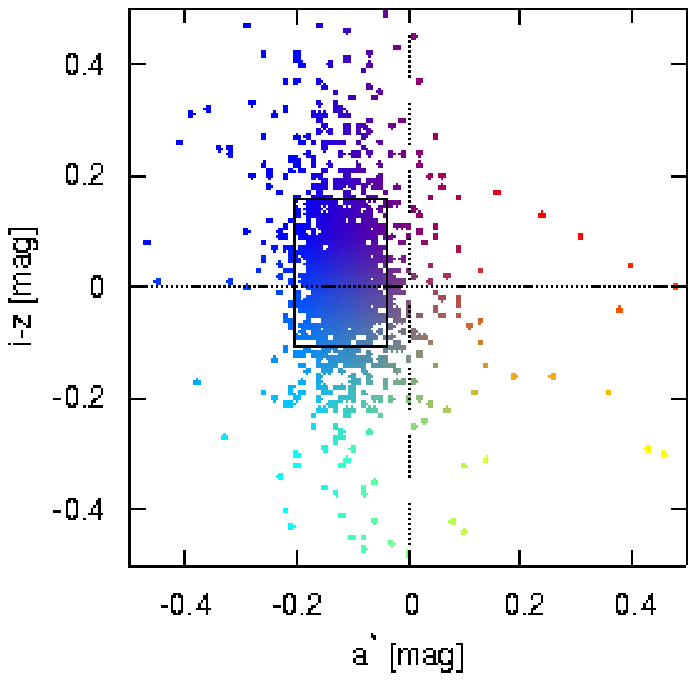}\,\,\includegraphics{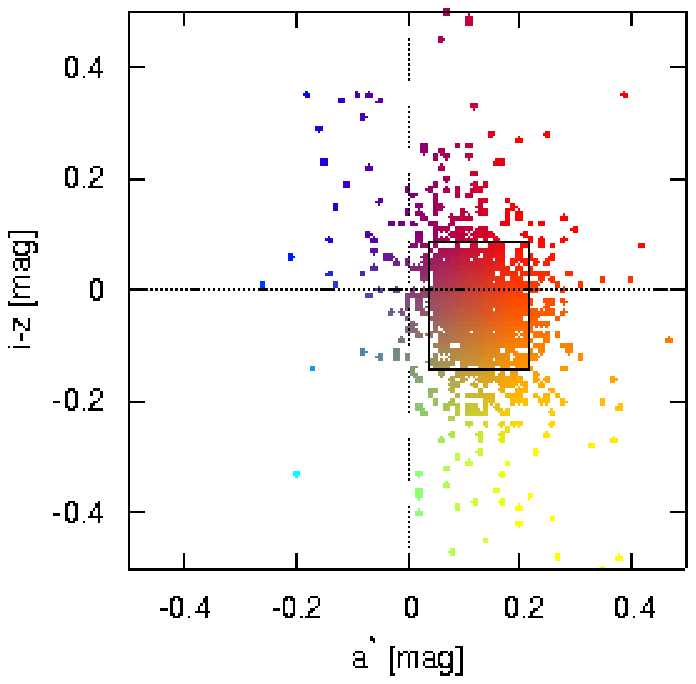}}\\
\end{center}
\caption{\label{img:sdss}Dependence of the color indexes $a^{\star}$ and $i-z$ (from the Sloan Digital Sky Survey Moving Object Catalog 4) for a C-type family Themis and S-type family Eunomia. The family corresponds to a compact structure in this parameter space marked by a rectangle. Note a qualitative difference between C- and S-types asteroids.}
\end{figure}

The HCM method selects a group of objects that are separated in the proper element space by less than a selected distance. However, not all of these objects are actually real members of the collisionally-born asteroid family. A~fraction of objects has orbital elements similar to typical elements of the asteroid family members only by a coincidence, the so-called interlopers. Interlopers can be identified (and removed), for example, by: 

\begin{itemize}
\item inspection of reflectance spectra, because they are usually of different taxonomic types than that of the family members, we use the SMASSII \citep{Bus2002} or Tholen taxonomy \citep{Tholen1984,Tholen1989};
\item inspection of colors based on the Sloan Digital Sky Survey Moving Object Catalog 4 \citep[SDSS MOC4,][]{Parker2008}, we used the color indexes $a^{\star}$ and $i-z$, which usually well define the core of the family (see examples for Themis and Eunomia families in Figure~\ref{img:sdss}), for each asteroid with available color indexes, we compared values $a^{\star}$ and $i-z$ to those that define the family;
\item inspection of albedos based on the WISE data \citep{Masiero2011};
\item constructing a diagram of the proper semi-major axis vs. the absolute magnitude (see Figure~\ref{img:a_H}), estimating the {\em V-shape} defined by the Yarkovsky semi-major axis drift and excluding outliers, i.e. relatively large asteroids outside the V-shape \citep[see][for the case of Eos family]{Vokrouhlicky2006}. We refer here the ($a_{\mathrm{p}}$, $H$) border of the family as the border of the V-shape;
\item constructing a size-frequency distribution (SFD) of the cluster, some asteroids can be too large to be created within the family and thus are believed to be interlopers \citep[see, e.g., numerical simulations by][who excluded the asteroid (490)~Veritas from the Veritas family]{Michel2011}.
\end{itemize}

These methods for family membership determination have one common characteristic -- we have to determine or choose a range for a quantity that defines the family members (range of spectra, sizes, or distance from the family center), which affects the number of objects we include in the family. Our criteria correspond to the fact that usually 99\% of the objects are within the ranges.

\subsection{New asteroid models}

From the DAMIT database, we adopt 96 models of asteroids that are, according to the HCM method, members of collisional families.

Currently, we have about 100 new asteroid models that have not yet been published. Here, we present new physical models of 31 asteroids from this sample that are identified as members of asteroid families by the HCM method (we choose only asteroids that belong to ten specific families for which we expect a reasonable amount of members, i.e. at least three). These convex shape models are derived by the lightcurve inversion method from combined dense and sparse photometry. The derivation process is similar to the one used in \citet{Hanus2013a}. The dense photometry was from two main sources:
(i)~the Uppsala Asteroid Photometric Catalogue \citep[UAPC\footnote{\texttt{http://asteroid.astro.helsinki.fi/}}, ][]{Lagerkvist1987, Piironen2001}, where lightcurves for about 1\,000 asteroids are stored, and
(ii)~the data from a group of individual observers provided by the Minor Planet Center in the Asteroid Lightcurve Data Exchange Format \citep[ALCDEF\footnote{\texttt{http://www.minorplanet.info/alcdef.html}},][]{Warner2009}.
The sparse-in-time photometry is downloaded from the AstDyS site (Asteroids -- Dynamic Site\footnote{\texttt{http://hamilton.dm.unipi.it/}}). We use data from the three most accurate observatories: USNO--Flagstaff station (IAU code 689), Roque de los Muchachos Observatory, La Palma (IAU code 950), and Catalina Sky Survey Observatory \citep[CSS for short, IAU code 703,][]{Larson2003}.

To increase the number of asteroid models for our study of asteroid families, we perform additional analysis of our previous results of the lightcurve inversion. For many asteroids, we are able to determine a unique rotational period, but get multiple pole solutions (typically 3--5) with similar ecliptic latitudes $\beta$, which is an important parameter. In \citet{Hanus2011}, we presented a reliability test, where we checked the physicality of derived solutions by the lightcurve inversion (i.e., if the shape model rotated around its axis with a maximum momentum of inertia). By computing models for all possible pole solutions and by checking their physicality, we remove the pole ambiguity for several asteroids, and thus determine their unique solutions (listed in Table~\ref{tab:models}). For other asteroids, the pole ambiguity remain and the models give us accurate period values and also rough estimates of ecliptic latitudes $\beta$ (if the biggest difference in latitudes of the models is $<50^{\circ}$). We call these models {\em partial} and present them in Table~\ref{tab:partials}. For the ecliptic latitude $\beta$, we use the mean value of all different models. We define parameter $\Delta\equiv|\beta_{\mathrm{max}}-\beta_{\mathrm{min}}|/2$ as being the estimated uncertainty of $\beta$, where $\beta_{\mathrm{max}}$ and $\beta_{\mathrm{min}}$ are the extremal values within all $\beta$. The threshold for partial models is $\Delta<25^{\circ}$. We present 31 new models and 24 partial models. References to the dense lightcurves used for the model determination are listed in Table~\ref{tab:references}. In Section~\ref{sec:simulation}, we compare the numbers of asteroids in four quadrants of the ($a_{\mathrm{p}}$, $\beta$) diagram (defined by the center of the family and the value $\beta=0^{\circ}$) with the same quantities based on the synthetic family population. The uncertainties in $\beta$ are rarely larger than 20$^{\circ}$, and the assignment to a specific quadrant is usually not questionable (only in 4 cases out of 136 the uncertainty interval lies in both quadrants, most of the asteroids have latitudes $|\beta|\gtrsim 30^{\circ}$), and thus give us useful information about the rotational properties in asteroid families. Partial models represent about 20\% of our sample of asteroid models.

The typical error for the orientation of the pole is (5--10$^{\circ}$)/$\cos \beta$ in longitude $\lambda$ and 5--20$^{\circ}$ in latitude $\beta$ (both uncertainties depend on the amount, timespan and quality of used photometry). Models based purely on dense photometry are typically derived from a large number ($\sim$30--50) of individual dense lightcurves observed during $\sim$5--10 apparitions, and thus the uncertainties of parameters of the rotational state correspond to lower values of the aforementioned range. On the other hand, models based on combined sparse-in-time data have, due to the poor photometric quality of the sparse data, the uncertainties larger (corresponding to the upper bound of the aforementioned range).

Models of asteroids (281)~Lucretia and (1188)~Gothlandia published by \citet{Hanus2013a} were recently determined also by \citet{Kryszczynska2013a} from partly different photometric data sets. Parameters of the rotational state for both models agree within their uncertainties.

The spin vector solution of asteroid (951)~Gaspra based on Galileo images obtained during the October 1991 flyby was already published by \citet{Davis1994}. Similarly, the solution of a Koronis-family member (243)~Ida based on Galileo images and photometric data was previously derived by \citet{Davies1994} and \citet{Binzel1993}. Here we present convex shape models for both these asteroids. Our derived pole orientations agree within only a few degrees with the previously published values (see Table~\ref{tab:families}), which again demonstrates the reliability of the lightcurve inversion method.

\subsection{Family members and interlopers}\label{sec:members}

We revise the family membership assignment by the HCM method according to the above-described criteria for interlopers or borderline cases. Interlopers are asteroids which do not clearly belong to the family, for example, they have different taxonomic types, incompatible albedos or are far from the ($a_{\mathrm{p}}$, $H$) border. On the other hand, borderline cases cannot be directly excluded from the family, their physical or orbital properties are just not typical in the context of other members (higher/lower albedos, close to the ($a_{\mathrm{p}}$, $H$) border). These asteroids are possible family members, but can be easily interlopers as well. In Table~\ref{tab:families}, last but one column, we show our revised membership classification of each object ({\em M} is a member, {\em I} an interloper and {\em B} a borderline case), the table also gives the rotational state of the asteroid (the ecliptic coordinates of the pole orientation $\lambda$ and $\beta$ and the period $P$), the semi-major axis $a$, the diameter $D$ and the albedo $p_{\mathrm{V}}$ from WISE \citep{Masiero2011}, the SMASS II \citep{Bus2002} and Tholen taxonomic types \citep{Tholen1984, Tholen1989}, and the reference to the model). 

Although we got for Vesta and Nysa/Polana families several members by the HCM method, we excluded these two families from our further study of spin states. Vesta family was created by a cratering event, and thus majority of fragments are rather small and beyond the capabilities of the model determination. Most of the models we currently have (recognized by the HCM method) are not compatible with the SFD of the Vesta family and thus are interlopers. On the other hand, Nysa/Polana family is a complex of two families (of different age and composition), thus should be treated individually. Additionally, we have only five member candidates for the whole complex, so even if we assign them to the subfamilies, the numbers would be too low to make any valid conclusions. 

In Table~\ref{tab:interlopers}, we list asteroids for which the HCM method suggested a membership to families Flora, Koronis, Eos, Eunomia, Phocaea and Alauda, but using the additional methods for the family membership determination described above, we identified them as interlopers or borderline cases.

\begin{figure*}
\begin{center}
\resizebox{\hsize}{!}{\includegraphics{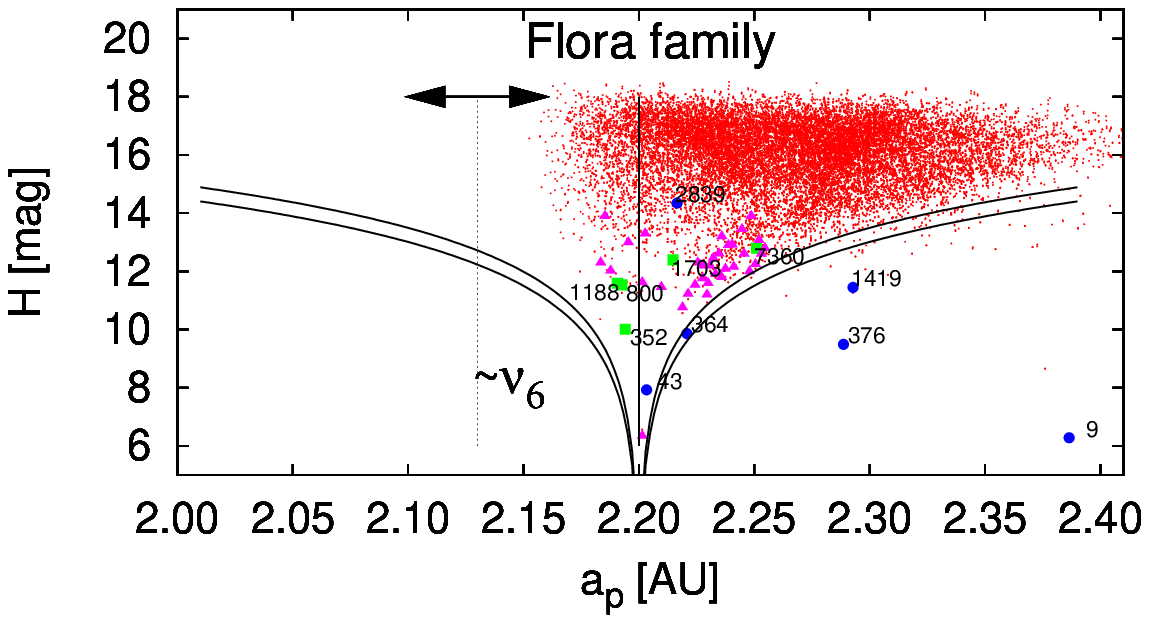}\includegraphics{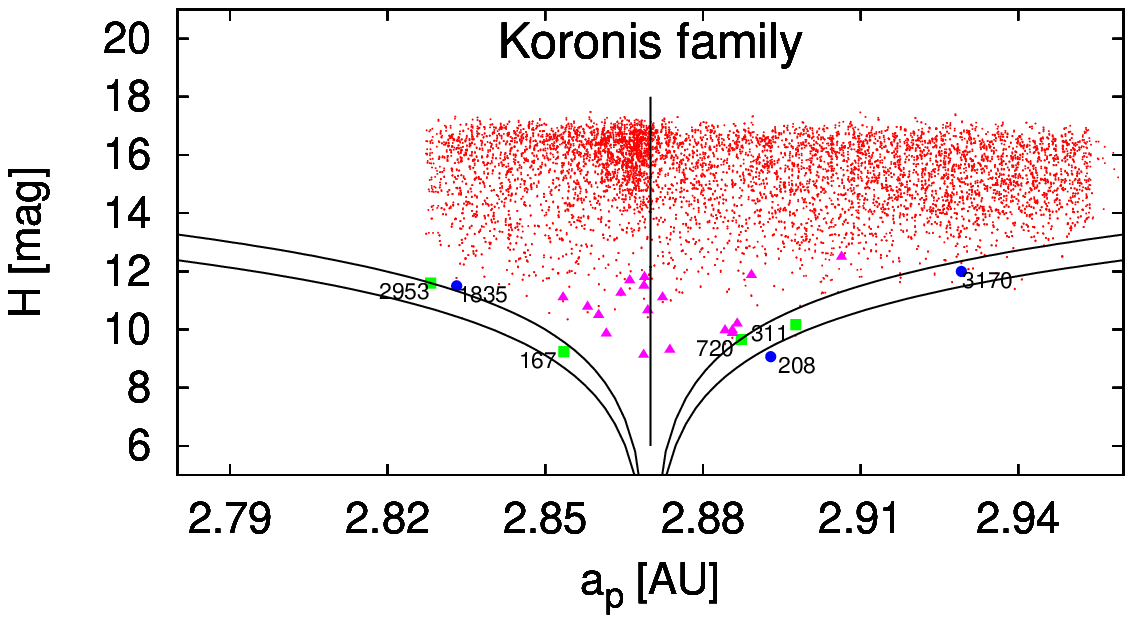}}\\
\resizebox{\hsize}{!}{\includegraphics{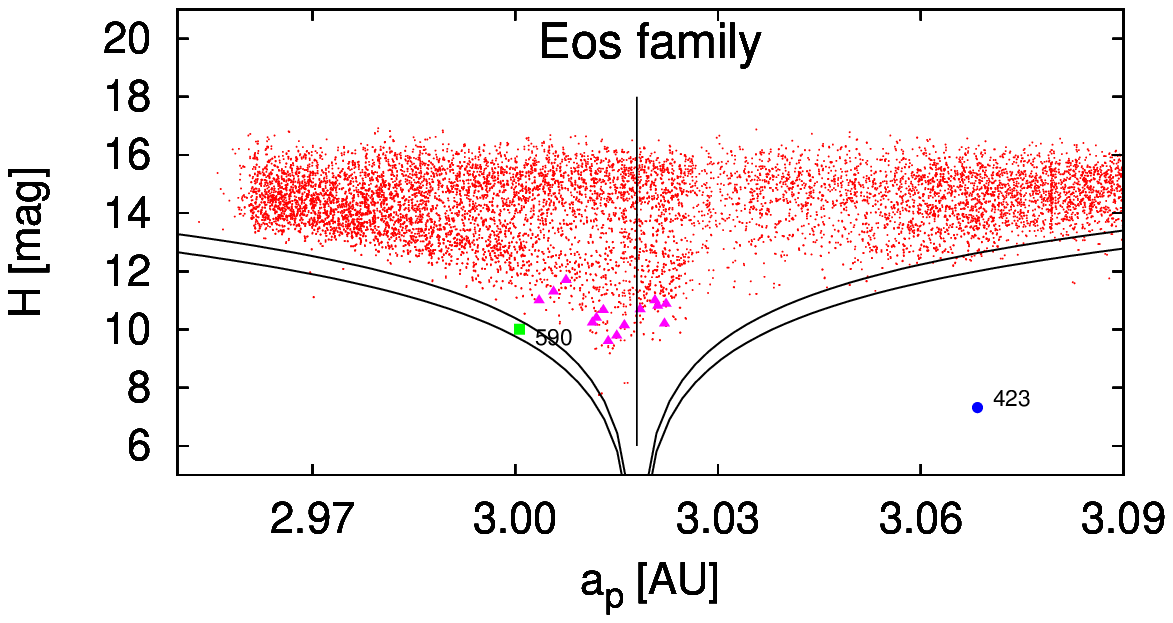}\includegraphics{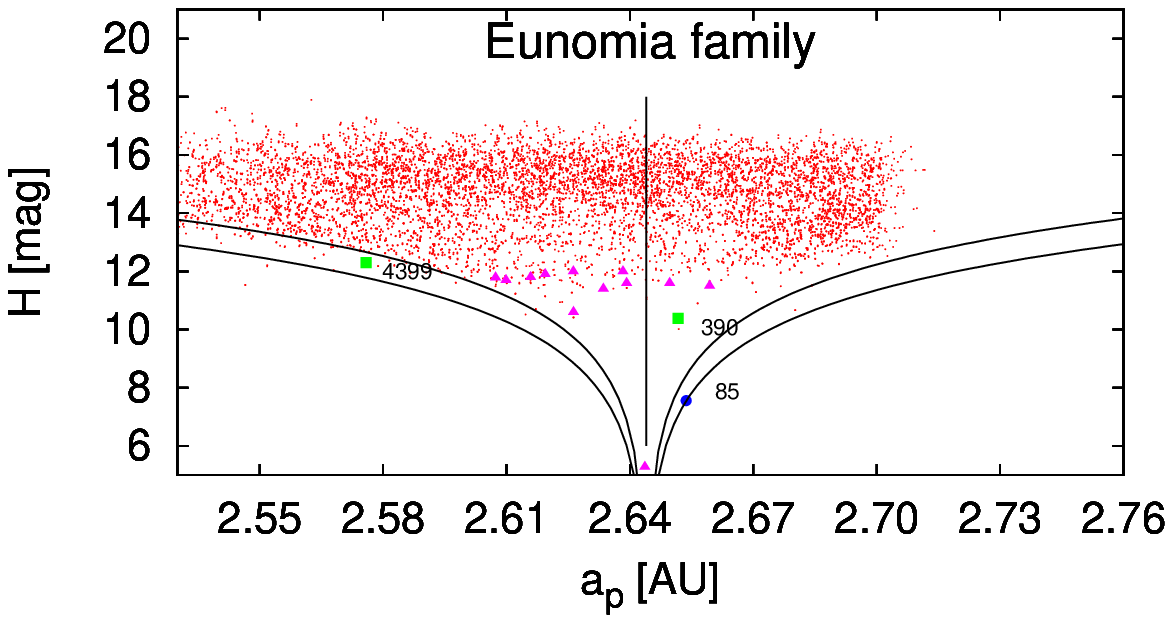}}\\
\resizebox{\hsize}{!}{\includegraphics{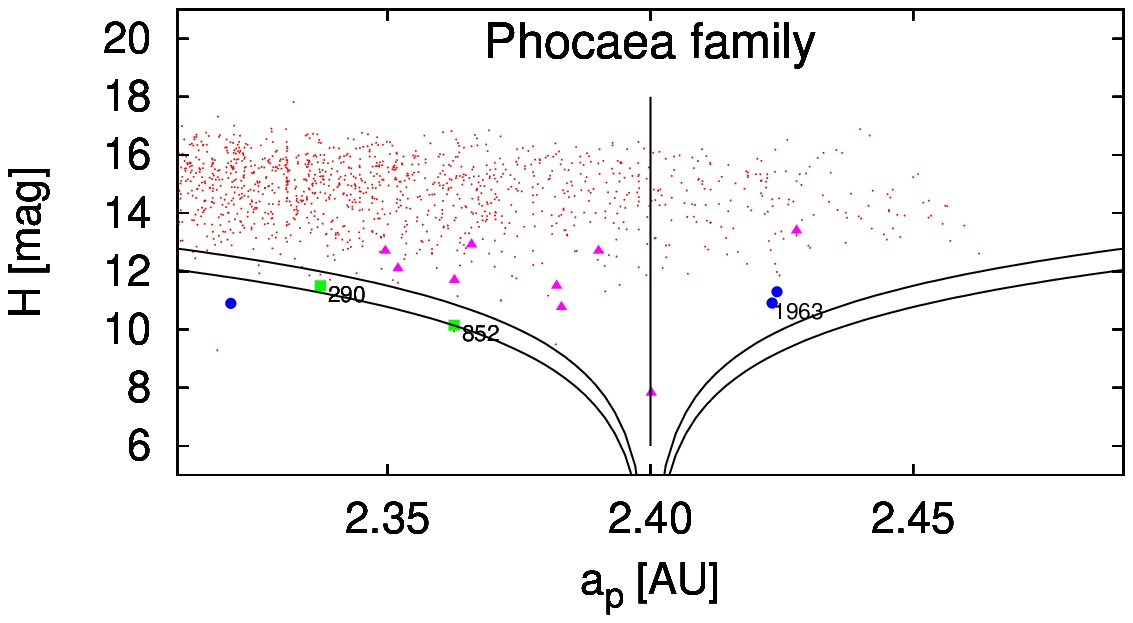}\includegraphics{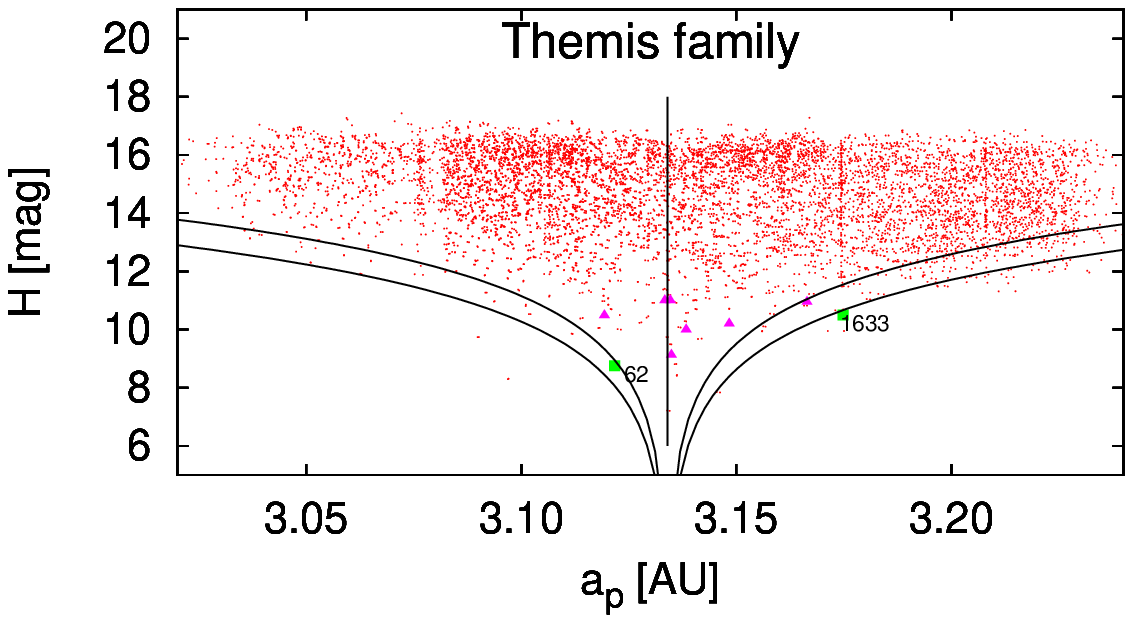}}\\
\resizebox{\hsize}{!}{\includegraphics{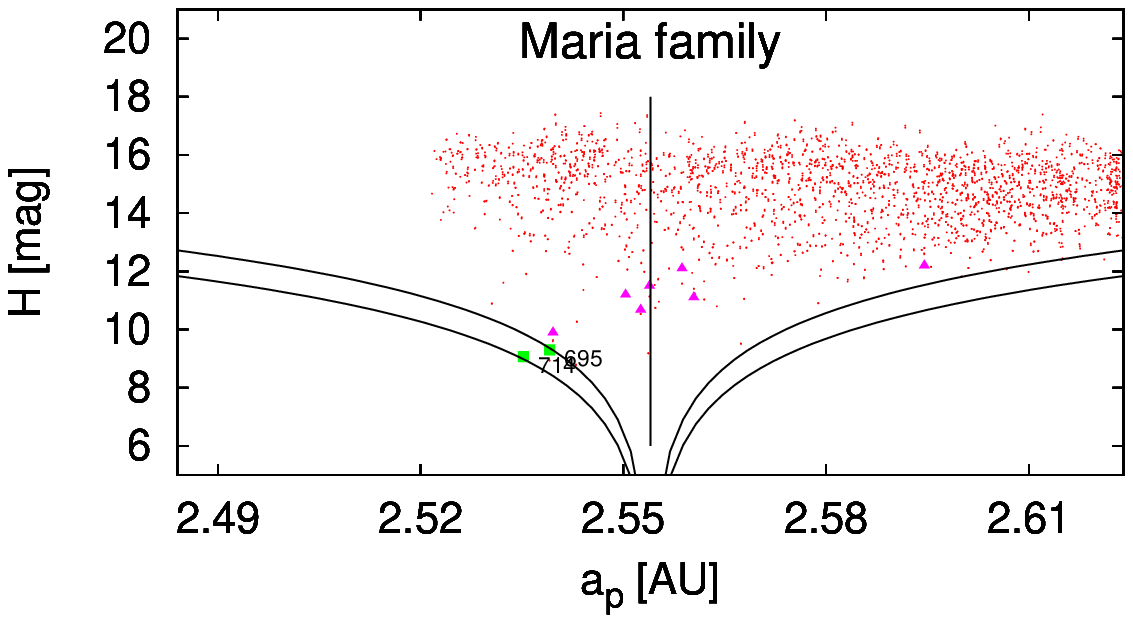}\includegraphics{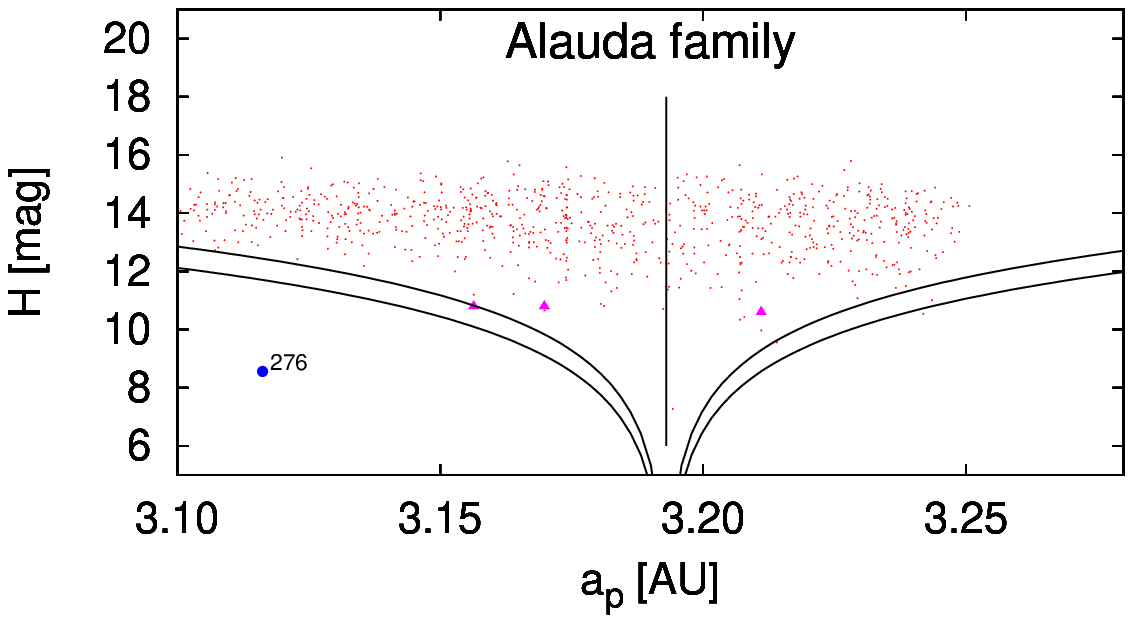}}\\
\end{center}
\caption{\label{img:V_shape}Dependence of the absolute magnitude $H$ on the proper semi-major axis $a_{\mathrm{p}}$ for the eight studies families: Flora, Koronis, Eos, Eunomia, Phocaea, Themis, Maria and Alauda with the likely positions of the family centers (vertical lines). We also plot the possible range of the ($a_{\mathrm{p}}$, $H$) borders (two thick lines) of each family for values of the parameter $C$ from \citet{Broz2013b} (different values correspond to a different initial extent of the family or different age and magnitude of the Yarkovsky semi-major axis drift.). The pink triangles represent the members from our sample (M), green circles borderline cases (B) and blue circles interlopers (I). Note that borderline cases and interlopers are identified by several methods including the position in the ($a_{\mathrm{p}}$, $H$) diagram, and thus could also lie close to the center of the family (e.g., in the case of the Flora family).}
\end{figure*}

In Figure~\ref{img:V_shape}, we show the ($a_{\mathrm{p}}$, $H$) diagrams for all eight studied families. We plot the adopted ($a_{\mathrm{p}}$, $H$) border \citep[from][]{Broz2013b} and label the members, borderline cases and interlopers by different colors.

Several asteroids in our sample belong to smaller and younger sub-clusters within the studied families (e.g, (832)~Karin in the Koronis family, (1270)~Datura in the Flora family or (2384)~Schulhof in the Eunomia family). These sub-clusters were likely created by secondary collisions. As a result, the spin states of asteroids in these sub-clusters were randomly reoriented. Because our combined orbital- and spin-evolution model (see Section~\ref{sec:simulation}) includes secondary collisions (reorientations), using asteroids from sub-clusters in the study of the spin-vector distribution is thus essential: asteroids from sub-clusters correspond to reoriented asteroids in our synthetic population.

\section{Observed spin vectors in families}\label{sec:analysis}

\begin{figure*}
\begin{center}
\resizebox{\hsize}{!}{\includegraphics{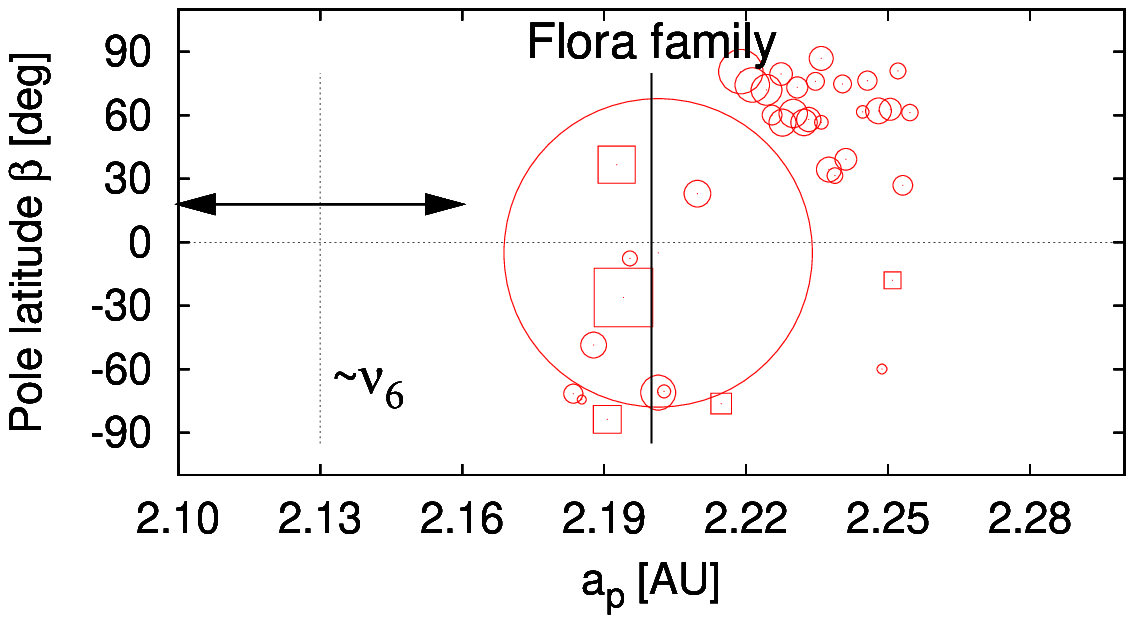}\includegraphics{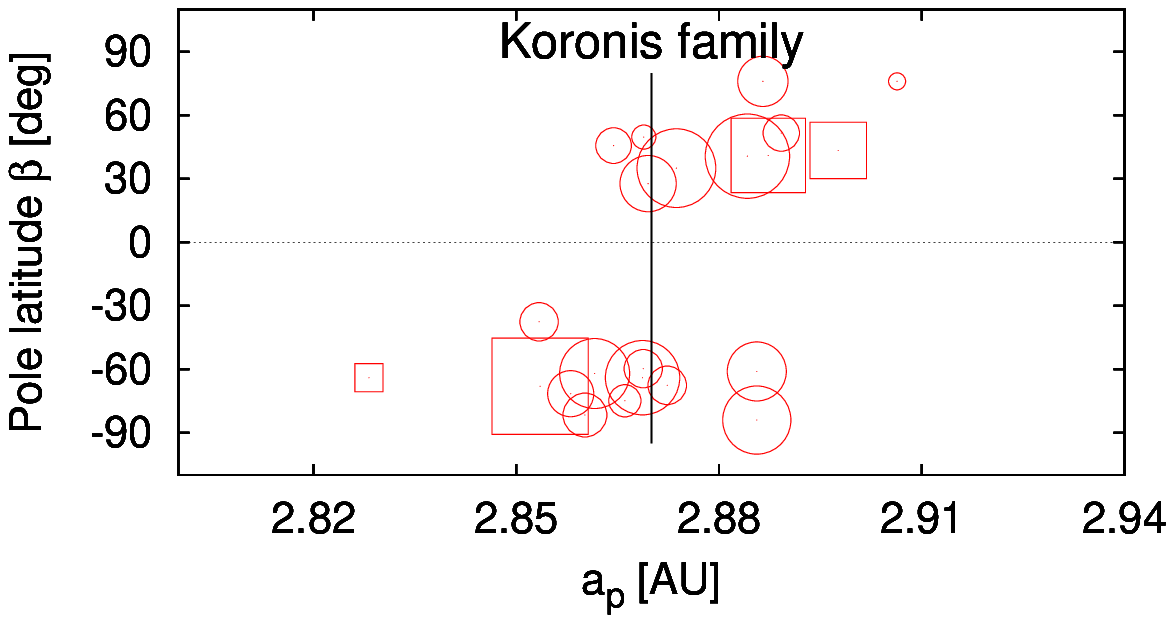}}\\
\resizebox{\hsize}{!}{\includegraphics{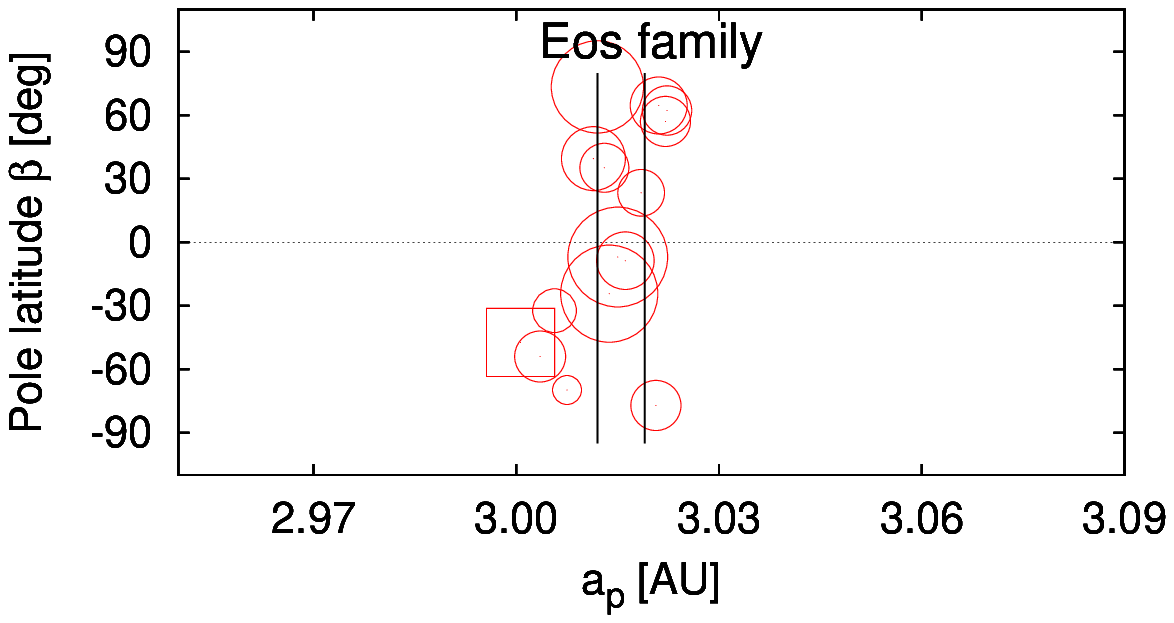}\includegraphics{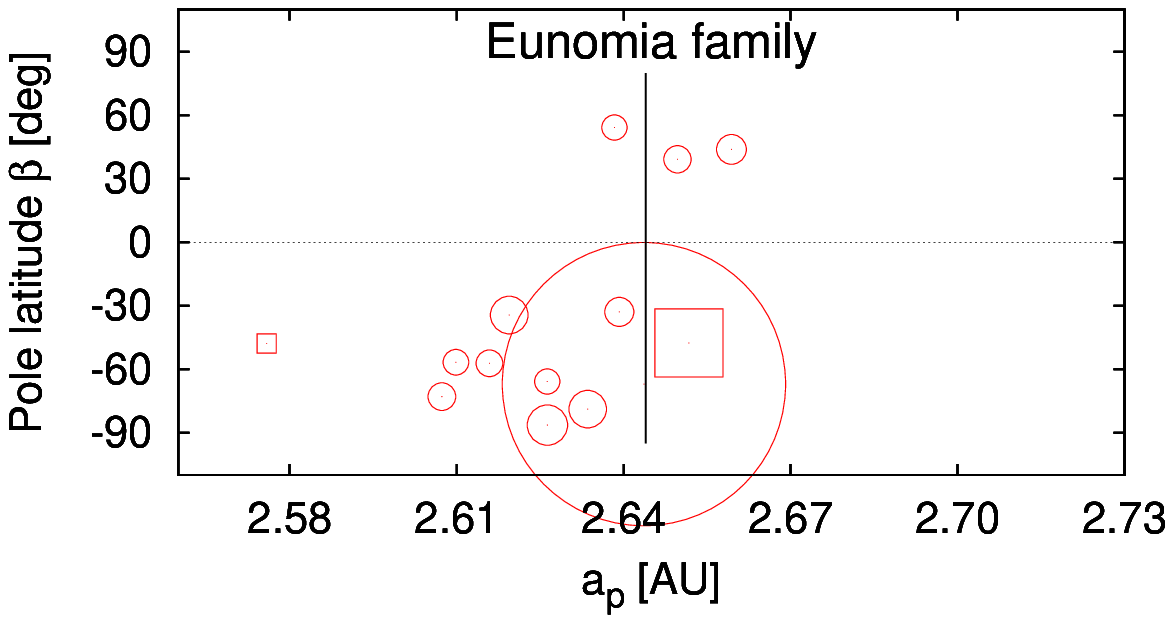}}\\
\resizebox{\hsize}{!}{\includegraphics{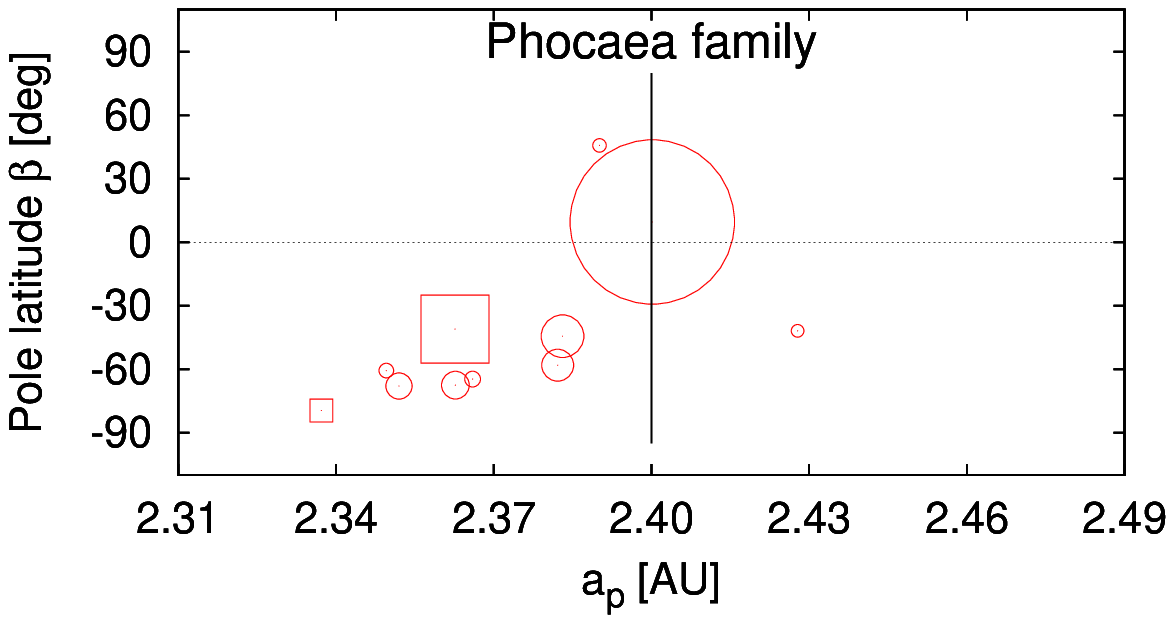}\includegraphics{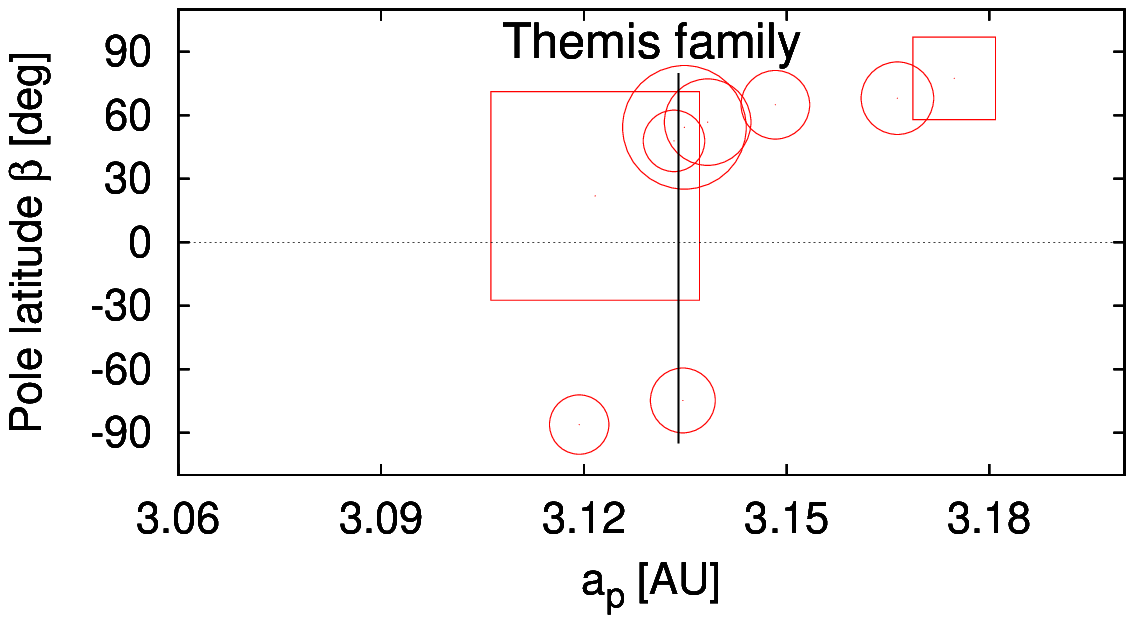}}\\
\resizebox{\hsize}{!}{\includegraphics{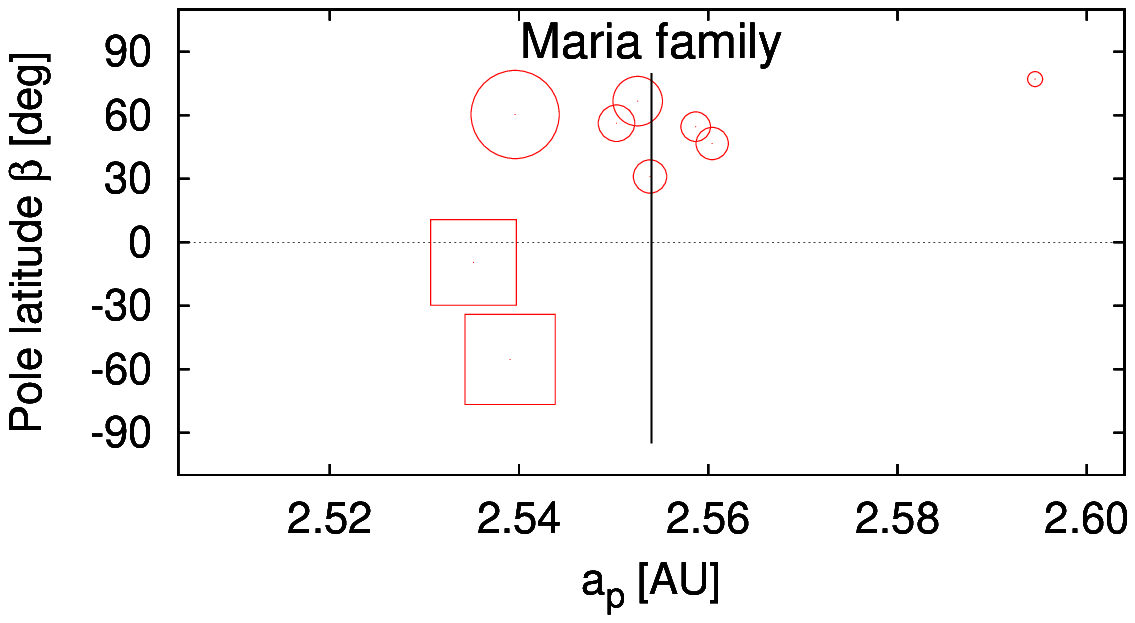}\includegraphics{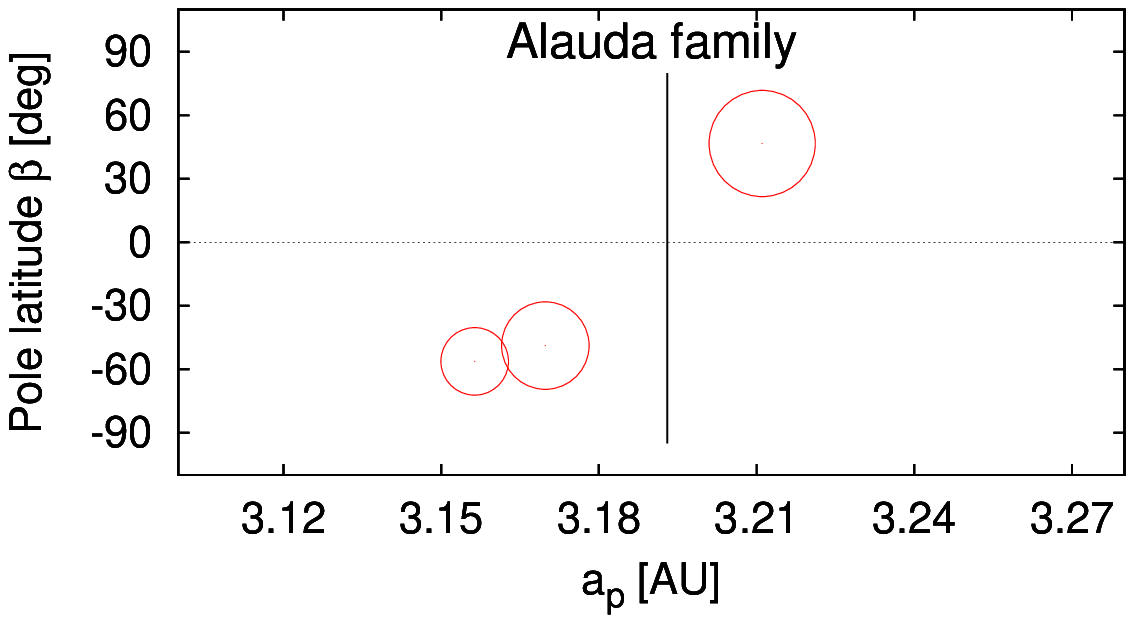}}\\
\end{center}
\caption{\label{img:families}Dependence of the pole latitude $\beta$ on the proper semi-major axis $a_{\mathrm{p}}$ for eight studied asteroid families: Flora, Koronis, Eos, Eunomia, Phocaea, Themis, Maria and Alauda. Family members are marked by circles and borderline cases by squares, which sizes are scaled proportionally to diameters (only the scale for (15)~Eunomia was decreased by half to fit the figure). The vertical lines correspond to the likely centers of the asteroid families, which uncertainties are usually $<0.01$ AU. The Eos family has an asymmetric V-shape (the ($a_{\mathrm{p}}$, $H$) border is asymmetric) which makes the center determination harder, so we marked two possible positions (one corresponds to the right ($a_{\mathrm{p}}$, $H$) border, the second to the left border). The uncertainties in $\beta$ are usually 5--20$^{\circ}$. In most cases, the value of $|\beta|\gtrsim 30^{\circ}$ and thus the quadrant to which the asteroid belongs (defined by the center of the family and the value $\beta=0^{\circ}$) is {\em not} changed.}
\end{figure*}

There are eight asteroid families for which we find at least three members (together with borderline cases) in our data set of asteroid models (after the family membership revision, labeled by {\em M} or {\em B} in the last column of Table~\ref{tab:families}) -- Flora (38 members), Koronis (23), Eos (16), Eunomia (14), Phocaea (11), Themis (9), Maria (9), and Alauda (3) families. Having the models and membership, we now can proceed to the discussion of the spin states in families in general (Section~\ref{sec:spin_state_general}), and for families Flora and Koronis (Sections~\ref{sec:flora},~\ref{sec:koronis}).

\subsection{Spin-vector orientations in individual families}\label{sec:spin_state_general}

In Figure~\ref{img:families}, we show the dependence of asteroid's pole latitudes in ecliptic coordinates (if there are two possible pole solutions for an asteroid, we take the first one in Table~\ref{tab:models}, because it corresponds to a formally better solution, additionally, latitudes for both ambiguous models are usually similar) on the semi-major axes. We mark the family members by circles and borderline cases by squares, which sizes are scaled proportionally to diameters to show also the dependence on the diameter. Vertical lines in Figure~\ref{img:families} correspond to the likely centers of the asteroid families, which we determine by constructing the V-shaped envelope of each family (we use all members of each family assigned by the HCM method, see Figure~\ref{img:a_H} and Figure~\ref{img:V_shape}). The Eos family has an asymmetric V-shape (the ($a_{\mathrm{p}}$, $H$) diagram), we compute centers for both wings of the V-shape individually. For the Flora family, we use only the right wing of the V-shape to derive the center, while the left one is strongly affected by the $\nu_6$ secular resonance.

In the study of spin-vector properties in families, we simply use the ecliptic coordinates for the pole orientation: ecliptic longitude $\lambda$ and latitude $\beta$. A formally better approach would be to use the coordinates bound to the orbital plane of the asteroid: orbital longitude $\lambda_{\mathrm{orb}}$ and latitude $\beta_{\mathrm{orb}}$. The orbital latitude can be then easily transformed to obliquity, which directly tells us, if the asteroid rotates in a prograde or retrograde sense. However, due to several reasons, we prefer the ecliptic coordinates: 
(i)~most of the asteroids have low inclinations and thus the difference between their ecliptic and orbital latitudes are only few degrees, the maximum differences for the families with higher inclination (Eos, Eunomia, Phocaea, Maria) are 20--30$^{\circ}$;
(ii)~the orbital coordinates of the pole direction cannot be computed for partial models, because we do not know the ecliptic longitude, these models represent about 20\% of our studied sample;
(iii)~the positions of the asteroids in the ($a_{\mathrm{p}}$, $\beta$) diagrams (i.e., to which quadrant they belong), namely if they have $\beta>0^{\circ}$ or $\beta<0^{\circ}$ are the sufficient information. Because most of the asteroids have latitudes larger than 30$^{\circ}$, their positions in the ($a_{\mathrm{p}}$, $\beta_{\mathrm{orb}}$) are similar (this is not true only for three asteroids out of 136); and
(iv)~we compare the ($a_{\mathrm{p}}$, $\beta$) diagrams (numbers of objects in the quadrants) between the observed and synthetic populations for ecliptic latitudes, so the consistency is assured.  

In general, we observe for all studied families similar trends:
(i) larger asteroids are situated in the proximity of the center of the family;
(ii) asteroids with $\beta>0^{\circ}$ are usually found to the right from the family center;
(iii) on the other hand, asteroids with $\beta<0^{\circ}$ to the left from the center;
(iv) majority of asteroids have large pole-ecliptic latitudes ($|\beta|\gtrsim30^{\circ}$); and finally
(v) some families have a statistically significant excess of asteroids with $\beta>0^{\circ}$ or $\beta<0^{\circ}$.

Case (i) is evident for families Flora, Eunomia, Phocaea, Themis or Maria. We have no large asteroids in the samples for the remaining families.

Cases (ii) and (iii) are present among all families with an exception of Eos, where all the asteroids are close to the (badly constrained) center. This phenomenon can be easily explained by the Yarkovsky drift, which can change asteroid's semi-major axes~$a$, namely it can increase $a$ of prograde rotators, and decrease $a$ of retrograde once. The magnitude of the Yarkovsky drift is dependent on the asteroid size, is negligible for asteroids with diameters $D\gtrsim$50 km (the case of Eos), and increases with decreasing diameter. For Flora, Eunomia, Phocaea or Maria family, we can see that the smallest asteroids in the sample ($D\sim$ 5--10 km) can be situated far from the family center, and we can also notice a trend of decreasing size with increasing distance from the center that probably corresponds to the magnitude of the Yarkovsky effect and the initial velocities $v_\mathrm{ini}(D)$ the objects gained after the break-up.

Observation (iv) is a result of the dynamical evolution of the asteroid's spin vector orientations dominated by the YORP effect, which increases the absolute value of the pole-ecliptic latitude \citep[see papers][where this effect is numerically investigated and compared with the observed anisotropic spin vector distribution of the sample of $\sim$300 MBAs]{Hanus2011, Hanus2013a}.

Case (v) concerns families Flora, Eunomia, Phocaea, Themis and Maria. The different number of asteroids with $\beta>0^{\circ}$ and $\beta<0^{\circ}$ among these families is statistically significant and cannot be coincidental. The obvious choice for an explanation are mean-motion or secular resonances. Indeed, the $\nu_6$ secular resonance removed many objects with $\beta>0^{\circ}$ from the Flora family (see Section~\ref{sec:flora} for a more thorough discussion), the 8:3 resonance with Jupiter truncated the Eunomia family, which resulted into the fact that there are no objects with $a_\mathrm{p}>2.70$~AU, and similarly, the 3:1 resonance with Jupiter affected the Maria family, for which we do not observe objects with smaller $a_\mathrm{p}$ than 2.52~AU. Near the Phocaea family at $a=2.50$~AU, the 3:1 resonance with Jupiter is situated. Due to the high inclination of objects in the Phocaea family ($I\sim24^{\circ}$), the resonance affects asteroids with $a_\mathrm{p}>2.40$~AU, which corresponds to the probable center of the family. The resonance removed a significant number of objects between 2.40~AU and 2.45~AU, and all objects with $a_\mathrm{p}$ larger.

The asymmetry of asteroids with $\beta>0^{\circ}$ and $\beta<0^{\circ}$ in the Themis family is caused by a selection effect: in the family, there are no objects with absolute magnitude $H<12$~mag (i.e., large asteroids) and $a_\mathrm{p}<3.10$~AU, on the other hand, with $a_\mathrm{p}>3.10$~AU, there are more than a hundred of such asteroids (see Figure~\ref{img:a_H}a). Our sample of asteroid models derived by the lightcurve inversion method is dominated by larger asteroids, and it is thus not surprising that we did not derive models for the Themis family asteroids with $a_\mathrm{p}<3.10$~AU.

The Flora and Koronis families are interesting also from other aspects, and thus are discussed in more detail in Sections~\ref{sec:flora} and~\ref{sec:koronis}.

\subsection{The Flora family}\label{sec:flora}

The Flora cluster is situated in the inner part of the main belt between 2.17--2.40 AU, its left part (with respect to the ($a_{\mathrm{p}}$, $H$) diagram) is strongly affected by the secular $\nu_6$ resonance with Saturn, which is demonstrated in Figure~\ref{img:a_H}b. The probable center of the family matches the position of asteroid (8)~Flora at $a=2.202$ AU. Because of the relative proximity to the Earth, more photometric measurements of smaller asteroids are available than for more distant families, and thus more models were derived. So far, we identified 38 models of asteroids that belong to the Flora family (together with borderline cases). 

The majority of asteroids within this family have $\beta>0^{\circ}$ ($\sim$68\%, due to small inclinations of the family members, majority of the objects with $\beta>0^{\circ}$ are definitely prograde rotators, because their obliquities are between 0$^\circ$ and 90$^\circ$) and lie to the right from the center of the family, confirming the presence of the Yarkovsky drift. Nine out of twelve asteroids with $\beta<0^{\circ}$ can be found in Figure~\ref{img:families} near or to the left from the center of the family. The exceptions are the borderline asteroids (1703)~Barry and (7360)~Moberg, and asteroid (7169)~Linda with $a_{\mathrm{p}}$ close to 2.25~AU (see Figure~\ref{img:families}). The borderline category already suggests that the two asteroids could be possible interlopers and their rotational state seems to support this statement. However, it is also possible that these asteroids have been reoriented by a non-catastrophic collisions. Rotational state of another borderline asteroid (800)~Kressmannia is also not in agreement with the Yarkovsky/YORP predictions, and thus it could be an interloper (or reoriented). The asteroid (7169)~Linda classified as member could still be an interloper, which was not detected by our methods for interloper removal, or could be recently reoriented by a non-catastrophic collision (the typical timescale for a reorientation \citep[][see~Eq.~\ref{eq:reorientation}]{Farinella1998} of this 4km-sized asteroid with rotational period $P=27.9$ h is $\tau_{\rm reor} \sim 500$~Myr, which is comparable with the age of the family). The depopulation of poles close to the ecliptic plane is also clearly visible.

The $\nu_6$ resonance to the left from the center of the family creates an excess of retrograde rotators not only among the family, but also among the whole main belt population if we use the currently available sample of asteroid models (there are $\sim$300 asteroid models in DAMIT database, in the Flora family, there are 14 more asteroids with $\beta>0^{\circ}$ than with $\beta<0^{\circ}$ (corresponds to the prograde excess), which corresponds to about 6\% of the whole sample, this bias needs to be taken into consideration, for example, in the study of rotational properties among MBAs). 

The missing asteroids with $\beta<0^{\circ}$ were delivered by this resonance to the orbits crossing the orbits of terrestrial planets and are responsible, for example, for the retrograde excess of the NEAs \citep{LaSpina2004}: the $\nu_6$ resonance contributes to the NEA population only by retrograde rotators, other major mean-motion resonances, such as the 3:1 resonance with Jupiter, deliver both prograde and retrograde rotators in a similar amount.

We did not observe a prograde group of asteroids with similar pole-ecliptic longitudes in the Flora family (i.e., a direct analog of the Slivan state in the Koronis family) that was proposed by \citet{Kryszczynska2013a}. Although \citet{Kryszczynska2013a} claims that Slivan states are likely observed among the Flora family, no corresponding clustering of poles of the prograde rotators is shown, particularly of ecliptic longitudes. We believe that the term {\em Slivan state} was used incorrectly there.

\subsection{The Koronis family}\label{sec:koronis}

The Koronis family is located in the middle main belt between 2.83--2.95~AU with the center at $a=2.874$~AU. We identified 23~members (together with borderline cases) with determined shape models.

The concept given by the Yarkovsky and YORP predictions work also among the Koronis family (asteroids with $\beta<0^{\circ}$ lie to the left from the family center, asteroids with $\beta>0^{\circ}$ to the right, see Figure~\ref{img:families}). In addition to that, \citet{Slivan2002} and \citet{Slivan2003} noticed that prograde rotators have also clustered pole longitudes. These asteroids were trapped in a secular spin-orbital resonance $s_6$ and are referred as being in Slivan states \citep{Vokrouhlicky2003}. Several asteroids were later recognized as being incompatible with the Slivan states, such as (832)~Karin and (263)~Dresda by \citet{Slivan2012}. 
Asteroid (832)~Karin is the largest member of a young \citep[$\sim$5.8 Myr,][]{Nesvorny2004} collisional family that is confined within the larger Koronis family. The spin state of (832)~Karin was thus likely affected during this catastrophic event and changed to a random state. 
Asteroid (263)~Dresda could be randomly reoriented by a non-catastrophic collision that is likely to happen for at least a few of 27 asteroids in the Koronis cluster with known spin state solutions, or its initial rotational state and shape did not allow a capture in the resonance. All four borderline asteroids have rotational states in agreement with the Yarkovsky/YORP concept which may support their membership to the Koronis cluster. On the other hand, rotational states of asteroids (277)~Elvira and (321)~Florentina do not match the expected values, and thus could be again interlopers or affected by reorientations.

Being trapped in the spin-orbital resonance does not necessarily mean that the asteroid is a member of the Koronis family, it rather indicates that its initial orbital position, the rotational state and the shape were favorable for being trapped in the resonance. For example, asteroids (311)~Claudia, (720)~Bohlinia, (1835)~Gajdariya and (3170)~Dzhanibekov have expected rotational states but are either rejected from the Koronis family or classified as borderline cases by our membership revision.

\section{Long-term evolution of spin vectors in asteroid families}\label{sec:simulation}

Here we present a comparison of the observed spin-vector orientations in several asteroid families with a numerical model of the temporal spin-vector evolutions. We use a {\it combined\/} orbital- and spin-evolution model, which was described in detail in \citet{Broz2011}. We need to account for the fact that the Yarkovsky semi-major axis drift sensitively depends on the orientation of the spin axis, which is in turn affected by the YORP effect and non-disruptive collisions. This model includes the following processes, which are briefly described in the text:
  (i)~impact disruption;
 (ii)~gravitational perturbations of planets;
(iii)~the Yarkovsky effect;
 (iv)~the YORP effect;
  (v)~collisions and spin-axis reorientations; and
 (vi)~mass shedding.

\paragraph{Impact disruption}

To obtain initial conditions for the family just after the breakup event we use a very simple model of an isotropic ejection of fragments from
the work of \citet{Farinella1994}. The distribution of velocities "at infinity" follows the function
\begin{equation}
{\rm d}N(v) {\rm d} v = C' v (v^2 + v_{\rm esc}^2)^{-(\alpha+1)/2} {\rm d} v\,,\label{dN_v}
\end{equation}
with the exponent $\alpha$ being a~free parameter, $C'$ a normalization constant and $v_{\rm esc}$ the escape velocity from the parent body,
which is determined by its size~$D_{\rm PB}$ and mean density~$\rho_{\rm PB}$ as
$v_{\rm esc} = \sqrt{(2/3) \pi G \rho_{\rm PB}}\, D_{\rm PB}\,.$
The distribution is usually cut at a selected maximum allowed velocity $v_{\rm max}$ to prevent outliers. The initial velocities $|v|$ of individual bodies are generated by a straightforward Monte--Carlo code and the orientations of the velocity vectors $\vec v$ in space are assigned randomly. We also assume that the velocity of fragments is independent of their size.

We must also select initial osculating eccentricity~$e_{\rm i}$ of the parent body, initial inclination~$i_{\rm i}$, as well as true anomaly~$f_{\rm imp}$ and argument of perihelion~$\omega_{\rm imp}$ at the time of impact disruption, which determine the initial shape of the synthetic family just after the disruption of the parent body.

\paragraph{Gravitational perturbations of planets}

Orbital integrations are performed using the SWIFT package
\citep{Levison1994}, slightly modified to include necessary
online digital filters and a second-order symplectic integrator
\citep{Laskar2001}. The second-order symplectic scheme allows
us to use a time-step up to $\Delta t = 91\,{\rm d}$.

Our simulations include perturbations by four outer planets,
with their masses, initial positions and velocities taken from the
JPL DE405 ephemeris \citep{Standish1997}.
We modify the initial conditions of the planets and asteroids
by a barycentric correction to partially account for the influence
of the terrestrial planets.
The absence of the terrestrial planets as perturbers
is a reasonable approximation in the middle and outer part
of the main belt (for orbits with $a > 2.5\,{\rm AU}$ and $e < 0.6$).%
\footnote{For the Flora family located in the inner belt we should account
for terrestrial planets directly, because of mean-motion resonances
with Mars, but we decided not do so, to speed-up the computation.
Anyway, the major perturbation we need to account for is the
$\nu_6$~secular resonance, which is indeed present in our model.}

Synthetic proper elements are computed as follows.
We first apply a Fourier filter to the (non-singular)
orbital elements in a moving window of 0.7 Myr (with steps
of 0.1 Myr) to eliminate all periods smaller than some threshold
(1.5 kyr in our case); we use a sequence of Kaiser windows
as in \citet{Quinn1991}.

The filtered signal, mean orbital elements, is then passed through
a frequency analysis code adapted from \citet{Sidlichovsky1996}
to obtain (planetary) forced and free terms in Fourier representation
of the orbital elements. The isolated free terms are what we use
as the proper orbital elements.

\paragraph{Yarkovsky effect} 

Both diurnal and seasonal components of the Yarkovsky accelerations
are computed directly in the N-body integrator. We use a theory of
\citet{Vokrouhlicky1998a} and \citet{Vokrouhlicky1999b}
for spherical objects \citep[but the magnitude of the acceleration does
not differ substantially for non-spherical shapes][]{Vokrouhlicky1998b}.
The implementation within the SWIFT integrator was described
in detail by \citet{Broz2006}.

\paragraph{YORP effect}\label{sec:yorp}

The evolution of the orientation of the spin axis and of the angular velocity is given by:
\begin{eqnarray}
\frac{{\rm d}\omega}{{\rm d} t} &=& c f_i(\epsilon)\,,\qquad i = 1 \dots 200\,,\label{eq:domega}\\
\frac{{\rm d}\epsilon}{{\rm d} t} &=& {c \frac{g_i(\epsilon)}{\omega}}\,,\label{eq:depsil}
\end{eqnarray}
where $f$- and $g$-functions describing the YORP effect for a set of 200 shapes were calculated numerically by \citet{Capek2004} with the effective radius $R_0 = 1\,{\rm km}$, the bulk density $\rho_0 = 2500\,{\rm kg}/{\rm m}^3$, located on a circular orbit with the semi-major axis $a_0 = 2.5\,{\rm AU}$. We assigned one of the artificial shapes (denoted by the index~$i$) to each individual asteroid from our sample.
The $f$- and $g$-functions were then scaled by the factor
\begin{equation}\label{cyorp}
c = c_{\rm YORP} \left(\frac{a}{a_0}\right)^{-2} \left(\frac{R}{R_0}\right)^{-2}\left(\frac{\rho_{\rm bulk}}{\rho_0}\right)^{-1}\,,
\end{equation}
where $a$, $R$, $\rho_{\rm bulk}$ denote the semi-major axis, the radius, and the density of the simulated body, respectively, and $c_{\rm YORP}$ is a free scaling parameter reflecting our uncertainty in the shape models and the magnitude of the YORP torque, which dependents on small-sized surface features \citep[even boulders,][]{Statler2009} and other simplifications in the modeling of the YORP torque. In \citet{Hanus2013a}, we constrained this parameter and find $c_{\rm YORP}=0.2$ to be the optimal value when comparing the results of the simulation with the observed latitude distribution of main belt asteroids. In our simulation, we used this value for $c_{\rm YORP}$.

The differential equations~(\ref{eq:domega}), (\ref{eq:depsil}) are integrated numerically by a simple Euler integrator. The usual time step is $\Delta t = 1000\,{\rm yr}$.

\paragraph{Collisions and spin-axis reorientations}\label{sec:reorientation}

We neglect the effect of disruptive collisions because we do not want to loose objects during the simulation, but we include spin axis reorientations caused by collisions. We use an estimate of the time scale by \citet{Farinella1998}.
\begin{equation}\label{eq:reorientation}
\tau_{\rm reor} = B \left(\frac{\omega}{\omega_0}\right)^{\beta_1} \left(\frac{D}{D_0}\right)^{\beta_2}\,,
\end{equation}
where
$B = 84.5\,{\rm kyr}$,
$\beta_1 = 5/6$,
$\beta_2 = 4/3$,
$D_0 = 2\,{\rm m}$ and
$\omega_0$ corresponds to period $P = 5$~hours.
These values are characteristic for the main belt.

\paragraph{Mass shedding}

If the angular velocity approaches a critical value
\begin{equation}
\omega_{\rm crit} = \sqrt{\frac{4}{3} \pi G \rho_{\rm bulk}}\,,
\end{equation}
we assume a mass shedding event, so we keep the orientation of the spin axis and the sense of rotation, but we reset the orbital period~$P = {2\pi/\omega}$ to a random value from the interval $(2.5, 9)$~hours. We also change the assigned shape to a different one, since any change of shape may result in a different YORP effect.

\paragraph{Synthetic Flora, Koronis and Eos families}

\begin{figure*}
\begin{center}
\resizebox{\hsize}{!}{\includegraphics{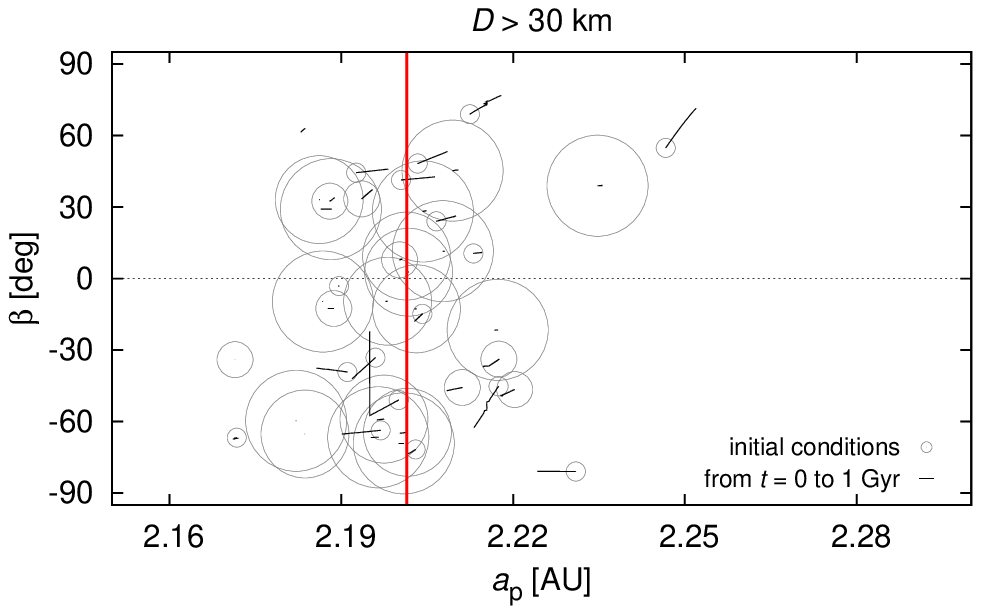}\includegraphics{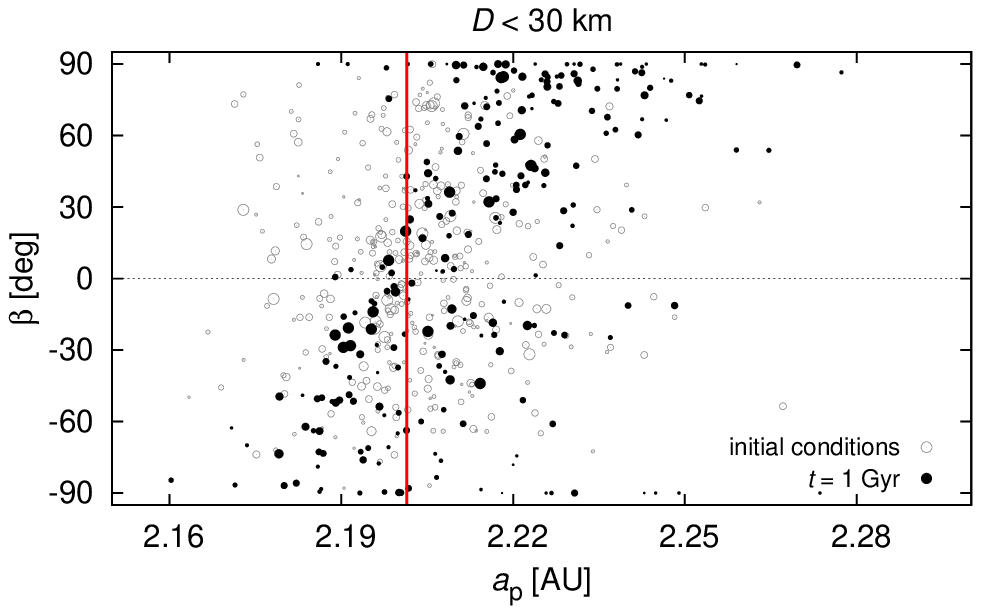}}\\
\resizebox{\hsize}{!}{\includegraphics{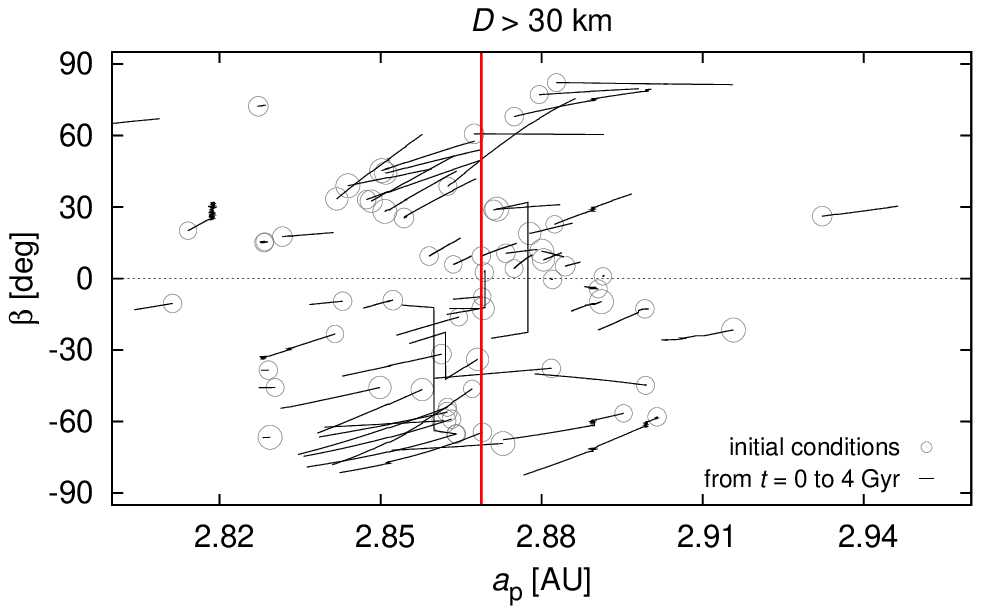}\includegraphics{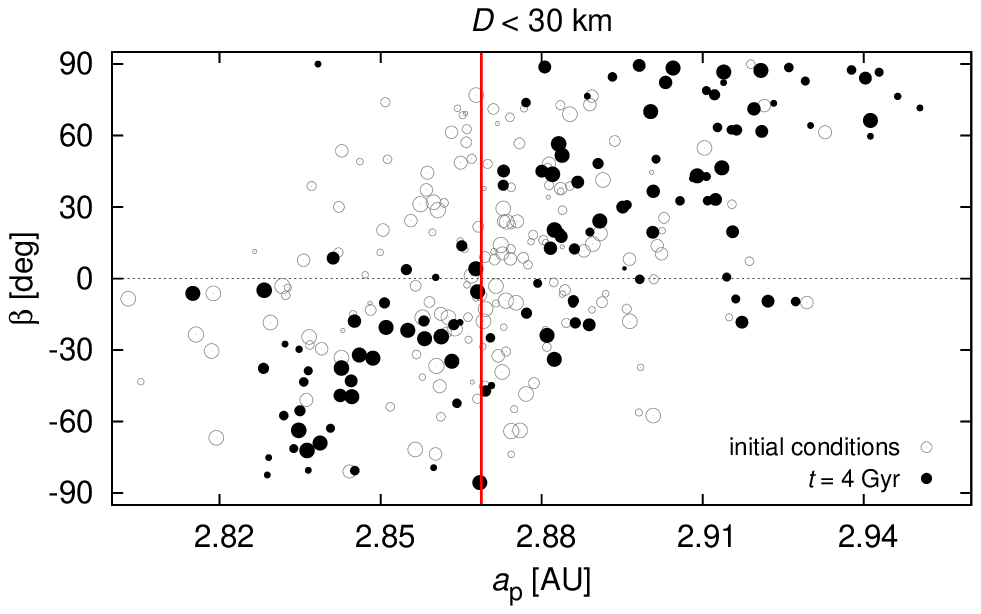}}\\
\resizebox{\hsize}{!}{\includegraphics{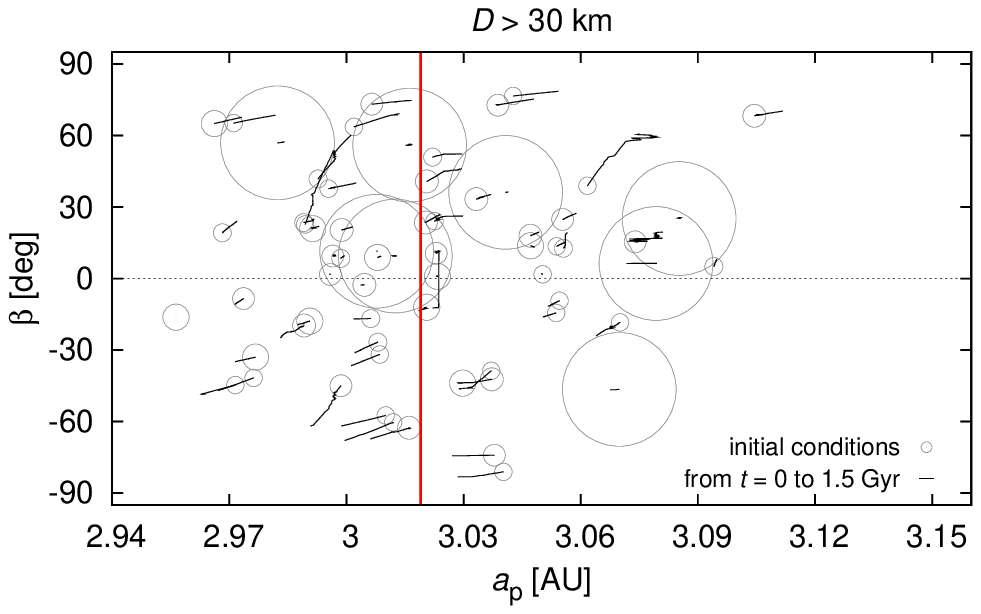}\includegraphics{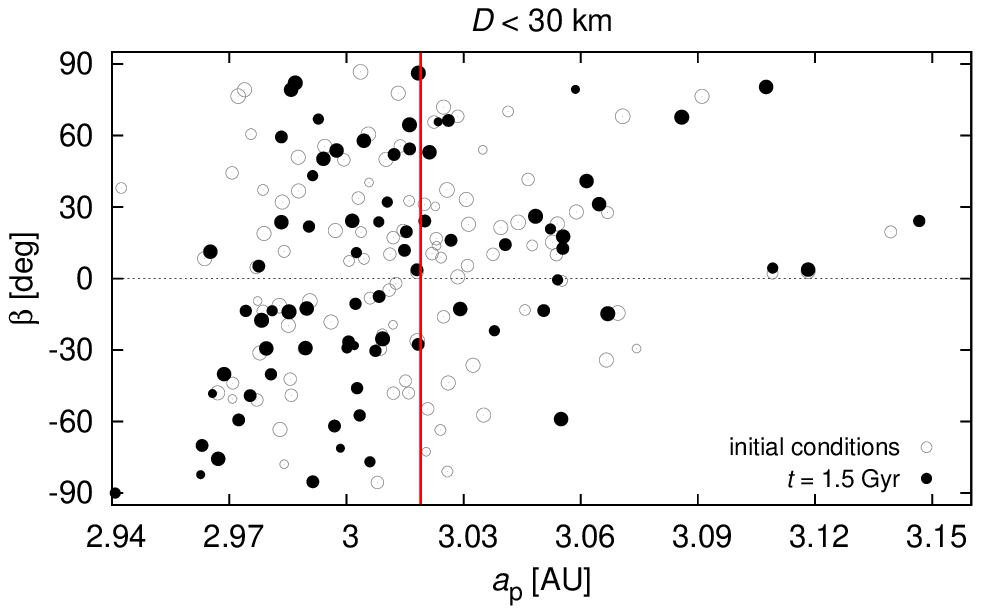}}\\
\end{center}
\caption{\label{img:flora_evolution}A simulation of the long-term evolution of the synthetic Flora (top), Koronis (middle) and Eos (bottom) families in the proper semi-major axis~$a_{\rm p}$ vs. the pole latitude~$\beta$ plane. Left: objects larger than $D > 30\,{\rm km}$, which almost do not evolve in $\beta$. Right: objects with $D \le 30\,{\rm km}$, with the initial conditions denoted by empty circles and an evolved state at 1 Gyr denoted by full circles. The sizes of symbols correspond to the actual diameters~$D$.
The initial conditions for Flora correspond to an isotropic size-independent velocity field with $\alpha = 3.25$ and $v_{\rm esc} = 95\,{\rm m}\,{\rm s}^{-1}$, and a uniform distribution of poles (i.e. $\sin\beta$). 
We increase the number of objects 10~times compared to the observed members of the Flora (Koronis and Eos as well) family in order to improve statistics. We retain their size distribution, of course.
The objects in Flora family are discarded from these plots when they left the family region (eccentricity $e_{\rm p} = 0.1\hbox{ to }0.18$, inclination $\sin I_{\rm p} = 0.05\hbox{ to }0.13$), because they are affected by strong mean-motion or secular resonances ($\nu_6$ in this case).
Thermal parameters were set as follows:
the bulk density $\rho_\mathrm{bulk} = 2500\,{\rm kg}\,{\rm m}^{-3}$,
the surface density $\rho_{\rm surf} = 1500\,{\rm kg}\,{\rm m}^{-3}$,
the thermal conductivity $K = 0.001\,{\rm W}\,{\rm m}^{-1}\,{\rm K}^{-1}$,
the thermal capacity $C_{\mathrm{t}} = 680\,{\rm J}\,{\rm kg}^{-1}$,
the Bond albedo $A = 0.1$ and
the infrared emissivity $\epsilon = 0.9$.
The time step for the orbital integration is ${\rm d} t = 91\,{\rm days}$ and~${\rm d} t_{\rm spin} = 10^3\,{\rm yr}$ for the (parallel) spin integration.
The parameters for Koronis and Eos are chosen similarly, only for Koronis, we use $v_{\rm esc} = 100\,{\rm m}\,{\rm s}^{-1}$, and for Eos $v_{\rm esc} = 225\,{\rm m}\,{\rm s}^{-1}$ and $\rho_{\rm surf} = 2500\,{\rm kg}\,{\rm m}^{-3}$. 
}
\end{figure*}

In Figure~\ref{img:flora_evolution} (top panel), we show a long-term evolution of the synthetic Flora family in the proper semi-major axis~$a_{\rm p}$ vs. the pole latitude~$\beta$ plane for objects larger and smaller than $30\,{\rm km}$. The values of the model parameters are listed in the figure caption. Larger asteroids do not evolve significantly and remain close to their initial positions. On the other hand, smaller asteroids ($D<30\,{\rm km}$) are strongly affected by the Yarkovsky and YORP effects: They drift in the semi-major axis, differently for prograde and retrograde rotators, and their pole orientations become mostly perpendicular to their orbits (corresponds to the proximity of the ecliptic plane for small inclinations). After the simulation at $t=1$~Gyr, we observe a deficiency of asteroids with $\beta>0^{\circ}$ to the left from the family center and a deficiency of asteroids with $\beta<0^{\circ}$ to the right from the family center. 

The asymmetry of the synthetic Flora family with respect to its center (red vertical line in the Figure~\ref{img:flora_evolution}) caused by the secular $\nu_6$ resonance is obvious. The down-right quadrant ($\beta<0^{\circ}$, $a_{\rm p}>2.202$~AU) still contains for $t=1$~Gyr many objects, because for some of them the evolution in $\beta$ and $a_{\rm p}$ is rather small, and other were delivered to this quadrant by collisional reorientations.    

The appearance of the evolved proper semi-major axis~$a_{\rm p}$ vs. the pole latitude~$\beta$ diagrams for Koronis and Eos families are qualitatively similar to the one of the Flora family. Because the asteroid samples for Koronis and Eos families are dominated by intermediate-sized asteroids ($D\sim20-50$~km), the evolution in $a_{\rm p}$ and $\beta$ is on average slower than in the Flora family. We show the state of the simulation for Koronis family in 4 Gyr and for Eos in 1.5 Gyr (based on the expected ages). The Eos family thus seems less evolved than Koronis family.  

We also check the distributions of the proper eccentricities and inclinations of the synthetic Flora/Koronis/Eos objects if they (at least roughly) correspond to the observed family. However, the number of objects to compare is rather low, and seems insufficient for a detailed comparison of distributions in 3D space of proper elements ($a_{\rm p}$, $e_{\rm p}$, $\sin I_{\rm p}$).

\paragraph{Ages of Flora, Koronis and Eos families}

\begin{figure}
\begin{center}
\resizebox{\hsize}{!}{\includegraphics{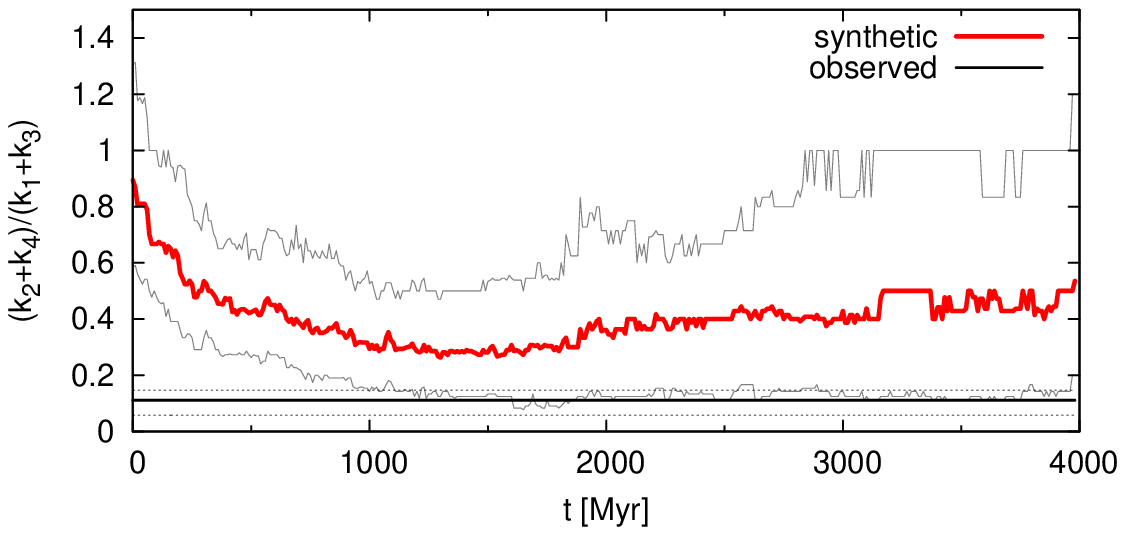}}\\
\resizebox{\hsize}{!}{\includegraphics{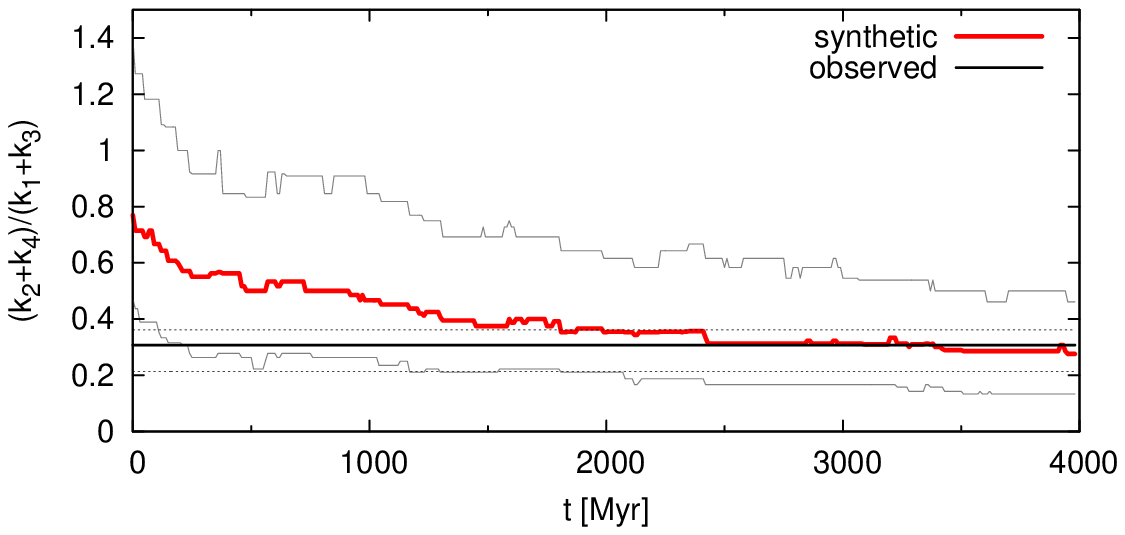}}\\
\resizebox{\hsize}{!}{\includegraphics{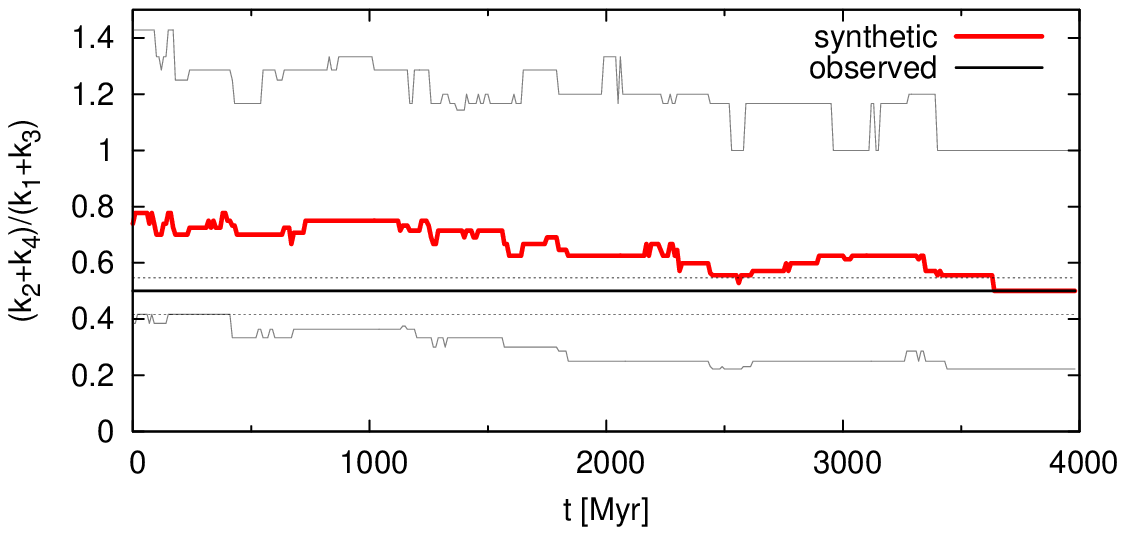}}\\
\end{center}
\caption{\label{img:ages}Time evolution of the metric $(k_2+k_4)/(k_1+k_3)$, where $k_i$ correspond to the numbers of synthetic objects in quadrants $i$ ($i=1, 2, 3, 4$) that are defined by the center of the family and value $\beta=0^{\circ}$, for synthetic Flora, Koronis and Eos families (red lines). The spread corresponds to 100 different selections of objects (we simulate 10 times more objects to reach a better statistics), the upper curve denotes the 90\% quantile and the bottom 10\%. Thick horizontal line is the observed ratio $(k_2+k_4)/(k_1+k_3)$ with the uncertainty interval.}
\end{figure}

To quantitatively compare the simulation of the long-term evolution of the synthetic families in the proper semi-major axis~$a_{\rm p}$ vs. the pole latitude~$\beta$ plane with the observation, we construct the following metric: we divide the ($a_{\rm p}$, $\beta$) plane into four quadrants defined by the center of the family and value $\beta=0^{\circ}$ and compute the ratio $(k_2+k_4)/(k_1+k_3)$, where $k_i$ correspond to the numbers of synthetic objects in quadrants $i$ ($i=1, 2, 3, 4$). In Figure~\ref{img:ages}, we show the evolution of the metric $(k_2+k_4)/(k_1+k_3)$ during the simulation of families Flora, Koronis and Eos for all synthetic objects with $D<30$~km, and the value of the same metric for the observed population for comparison.

For the Koronis family (middle panel), the synthetic ratio reaches the observed one after $t=2.5$~Gyr and remains similar until the end of the simulation at $t=4$~Gyr. \citet{Bottke2001} published the age $t=(2.5\pm1.0)$~Gyr for the Koronis family. Unfortunately, we cannot constrain the age of the Eos family from this simulation due to objects with the relatively small evolution in $a_{\rm p}$ and $\beta$. The fit for the Flora family is not ideal, the reason could be differences in the initial velocity field or the true anomaly $f_\mathrm{imp}$ of the impact. The best agreement is for the age $t=(1.0\pm0.5)$~Gyr, which is approximately in agreement with the dynamical age in \citet{Nesvorny2005}: $(1.5\pm0.5)$~ Gyr.

\section{Conclusions}

We identify 152 asteroids, for which we have convex shape models and simultaneously, the HCM method identifies them as members of ten collisional families. Due to a large number of expected interlopers in families Vesta and Nysa/Polana, we exclude these families from the study of the rotational properties. In the remaining sample of asteroids from eight families, we identify $\sim20$\% of objects that are interlopers or borderline cases (see Table~\ref{tab:interlopers}). We use several methods, described in Section~\ref{sec:membership}, for their identification. The borderline cases are still possible members of the families and thus are included in our study of the spin-vector distribution.

From the dependence of the asteroid's pole latitudes on the semi-major axes, plotted in Figure~\ref{img:families}, we can see fingerprints of families spreading in $a$ and spin axis evolution due to Yarkovsky and YORP effects: Asteroids with $\beta<0^{\circ}$ lie on the left side from the center of the family, and on the other hand, asteroids with $\beta>0^{\circ}$ on the right side. The asymmetry with respect to the family centers is in most cases caused by various resonances that cut the families, in the case of Themis family, a selection effect is responsible.

However, we do not observe a {\em perfect} agreement with the Yarkovsky and YORP effects predictions. A few (eight) individual objects that have incompatible rotational states could:
(i)~be incorrectly determined;
(ii)~be interlopers;
(iii)~have initial rotational states that cause only a small evolution in the ($a_{\rm p}$, $\beta$) space (i.e., they are close to their initial positions after the break-up); or
(iv)~be recently reoriented by collisional events.

In the case of the Flora family, significantly less asteroids with $\beta<0^{\circ}$ ($\sim28\%$) than with $\beta>0^{\circ}$ ($\sim72\%$) are present. The secular $\nu_6$ resonance is responsible for this strong deficit, because objects with $\beta<0^{\circ}$ are drifting towards this resonance and are subsequently removed from the family (they become part of the NEAs population where they create an excess of retrograde rotators). 

We do not find an analog of the Slivan states (observed in the Koronis family) among any other of the studied families.

We simulate a long-term evolution of the synthetic Flora, Koronis and Eos families (Figure~\ref{img:flora_evolution}) in the proper semi-major axis~$a_{\rm p}$ vs the pole latitude~$\beta$ plane and compare the results with the properties of observed asteroid families. We obtain a good qualitative agreement between the observed and synthetic spin-vector distributions. For all three families, we compute evolution of the number of objects in the four quadrants of the families in the ($a_{\rm p}$, $\beta$) diagram, and we estimate ages for families Flora $(1.0\pm0.5)$~Gyr and Koronis (2.5 to 4~Gyr) that are in agreement with previously published values. However, we do not estimate the age of the Eos family due to a small evolution of the objects in the ($a_{\rm p}$,$\beta$) diagram.

The uncertainties seem to be dominated by the observed quadrant ratios. We expect that increasing the sample size by a factor of 10 would decrease the relative uncertainty by a factor of about 3, which is a good motivation for further work on this subject.

\begin{acknowledgements}
The work of JH and JD has been supported by grants GACR P209/10/0537 and P209/12/0229 of the Czech Science Foundation, and the work of JD and MB by the Research Program MSM0021620860 of the Czech Ministry of Education. The work of MB has been also supported by grant GACR 13-013085 of the Czech Science Foundation.
\end{acknowledgements}

\bibliography{mybib}
\bibliographystyle{aa}

\onecolumn

\begin{longtable}{r@{\,\,\,}l rrrr D{.}{.}{6} ccccc}
\caption{\label{tab:models}List of new asteroid models derived from combined dense and sparse data or from sparse data alone.}\\
\hline 
\multicolumn{2}{c} {Asteroid} & \multicolumn{1}{c} {$\lambda_1$} & \multicolumn{1}{c} {$\beta_1$} & \multicolumn{1}{c} {$\lambda_2$} & \multicolumn{1}{c} {$\beta_2$} & \multicolumn{1}{c} {$P$} & $N_{\mathrm{lc}}$ & $N_{\mathrm{app}}$  & $N_{\mathrm{689}}$ & $N_{\mathrm{703}}$ & $N_{\mathrm{950}}$ \\
\multicolumn{2}{l} { } & [deg] & [deg] & [deg] & [deg] & \multicolumn{1}{c} {[hours]} &  &  &  &  & \\
\hline\hline

\endfirsthead
\caption{continued.}\\

\hline
\multicolumn{2}{c} {Asteroid} & \multicolumn{1}{c} {$\lambda_1$} & \multicolumn{1}{c} {$\beta_1$} & \multicolumn{1}{c} {$\lambda_2$} & \multicolumn{1}{c} {$\beta_2$} & \multicolumn{1}{c} {$P$} & $N_{\mathrm{lc}}$ & $N_{\mathrm{app}}$  & $N_{\mathrm{689}}$ & $N_{\mathrm{703}}$ & $N_{\mathrm{950}}$ \\
\multicolumn{2}{l} { } & [deg] & [deg] & [deg] & [deg] & \multicolumn{1}{c} {[hours]} &  &  &  &  & \\
\hline\hline
\endhead
\hline
\endfoot
243 & Ida & 259 & $-$66 & 74 & $-$61 & 4.633632 & 53 & 6 & 134 & 122 & 25 \\
364 & Isara & 282 & 44 & 86 & 42 & 9.15751 & 4 & 1 & 98 & 104 &  \\
540 & Rosamunde & 301 & 81 & 127 & 62 & 9.34779 & 3 & 1 & 135 & 83 &  \\
550 & Senta & 63 & $-$40 & 258 & $-$58 & 20.5726 & 9 & 1 & 151 & 85 &  \\
553 & Kundry & 197 & 73 & $-$1 & 64 & 12.6025 & 5 & 1 & 61 & 80 &  \\
621 & Werdandi & 247 & $-$86 & 66 & $-$77 & 11.77456 & 12 & 2 & 146 & 71 &  \\
936 & Kunigunde & 47 & 57 & 234 & 50 & 8.82653 &  &  & 154 & 88 &  \\
951 & Gaspra & 20 & 23 & 198 & 15 & 7.042027 & 71 & 4 & 117 & 89 & \\
1286 & Banachiewicza & 214 & 62 & 64 & 60 & 8.63043 &  &  & 81 & 51 &  \\
1353 & Maartje & 266 & 73 & 92 & 57 & 22.9926 &  &  & 154 & 139 &  \\
1378 & Leonce & 210 & $-$67 & 46 & $-$77 & 4.32527 &  &  & 89 & 113 &  \\
1423 & Jose & 78 & $-$82 &  &  & 12.3127 &  &  & 121 & 134 &  \\
1446 & Sillanpaa & 129 & 76 & 288 & 63 & 9.65855 & 2 & 1 & 76 & 73 &  \\
1464 & Armisticia & 194 & $-$54 & 35 & $-$69 & 7.46699 & 2 & 1 & 231 & 67 &  \\
1503 & Kuopio & 170 & $-$86 & 27 & $-$61 & 9.9586 &  &  & 116 & 68 &  \\
1527 & Malmquista & 274 & 80 &  &  & 14.0591 &  &  & 49 & 107 &  \\
1618 & Dawn & 39 & $-$60 & 215 & $-$51 & 43.219 &  &  & 93 & 91 &  \\
1633 & Chimay & 322 & 77 & 116 & 81 & 6.59064 & 2 & 1 & 127 & 83 &  \\
1691 & Oort & 45 & 68 & 223 & 58 & 10.2684 &  &  & 86 & 60 &  \\
1703 & Barry & 46 & $-$76 & 221 & $-$71 & 107.04 &  &  & 89 & 138 &  \\
1805 & Dirikis & 364 & 48 & 188 & 61 & 23.4543 &  &  & 117 & 91 &  \\
1835 & Gajdariya & 34 & 74 & 204 & 69 & 6.33768 &  &  & 66 & 86 &  \\
1987 & Kaplan & 356 & $-$58 & 233 & $-$89 & 9.45950 & 8 & 2 & 81 & 28 &  \\
2430 & Bruce Helin & 177 & $-$68 &  &  & 129.75 & 15 & 1 &  & 112 &  \\ 
3279 & Solon & 268 & $-$70 &  &  & 8.1043 & 3 & 1 &  & 137 &  \\
3492 & Petra-Pepi & 9 & $-$57 & 202 & $-$16 & 46.570 & 15 & 1 & 25 & 111 &  \\
4399 & Ashizuri & 266 & $-$48 & 45 & $-$61 & 2.830302 & 4 & 1 & 20 & 84 &  \\
4606 & Saheki & 44 & 59 & 222 & 68 & 4.97347 & 6 & 1 &  & 123 &  \\
6159 & 1991 YH & 266 & 67 & 62 & 67 & 10.6590 & 3 & 1 &  & 102 &  \\
6262 & Javid & 93 & 76 & 275 & 66 & 8.02054 & 3 & 1 &  & 106 &  \\
6403 & Steverin & 246 & 77 & 109 & 73 & 3.49119 & 2 & 1 &  & 74 &  \\
7043 & Godart & 73 & 62 & 235 & 80 & 8.4518 & 4 & 1 &  & 121 &  \\
7169 & Linda & 11 & $-$60 & 198 & $-$61 & 27.864 & 5 & 1 &  & 95 &  \\
\hline
\end{longtable}
\tablefoot{
For each asteroid, the table gives the ecliptic coordinates $\lambda_1$ and $\beta_1$ of the pole solution with the lowest $\chi^2$, the corresponding mirror solution $\lambda_2$ and $\beta_2$, the sidereal rotational period $P$, the number of dense lightcurves $N_{\mathrm{lc}}$ observed during $N_{\mathrm{app}}$ apparitions, and the number of sparse data points for the corresponding observatory: $N_{\mathrm{689}}$, $N_{\mathrm{703}}$ and $N_{\mathrm{950}}$. The uncertainty of the sidereal rotational period corresponds to the last decimal place of $P$ and of the pole direction to 5--10$^{\circ}$ if we have multi-apparition dense data or 10--20$^{\circ}$ if the model is based mainly on sparse data (i.e., only few dense lightcurves from 1--2 apparitions).
}

%
\begin{longtable}{r@{\,\,\,}l rrD{.}{.}{6}cccc}
\caption{\label{tab:partials}List of partial models derived from combined data sets.}\\
\hline 
 \multicolumn{2}{c} {Asteroid} & \multicolumn{1}{c} {$\beta$} & \multicolumn{1}{c} {$\Delta$} & \multicolumn{1}{c} {P} & N$_{\mathrm{lc}}$ & N$_{\mathrm{app}}$  & N$_{\mathrm{689}}$ & N$_{\mathrm{703}}$ \\
\multicolumn{2}{l} { } & [deg] & [deg] & \multicolumn{1}{c} {[hours]} &  &  &  &  \\
\hline\hline 

\endfirsthead
\caption{continued.}\\

\hline
 \multicolumn{2}{c} {Asteroid} & \multicolumn{1}{c} {$\beta$} & \multicolumn{1}{c} {$\Delta$} & \multicolumn{1}{c} {P} & N$_{\mathrm{lc}}$ & N$_{\mathrm{app}}$  & N$_{\mathrm{689}}$ & N$_{\mathrm{703}}$ \\
\multicolumn{2}{l} { } & [deg] & [deg] & \multicolumn{1}{c} {[hours]} &  &  &  &  \\
\hline\hline 
\endhead
\hline
\endfoot
391 & Ingeborg & $-$60 & 7 & 26.4145 & 24 & 2 & 141 & 96 \\
502 & Sigune & $-$44 & 3 & 10.92667 & 9 & 2 & 157 & 52 \\
616 & Elly & 67 & 23 & 5.29771 & 4 & 1 & 101 & 133 \\
1003 & Lilofee & 65 & 10 & 8.24991 & & & 107 & 83 \\
1160 & Illyria & 47 & 23 & 4.10295 & & & 96 & 100 \\
1192 & Prisma & $-$65 & 14 & 6.55836 & 5 & 1 & 44 & 43 \\
1276 & Ucclia & $-$49 & 22 & 4.90748 & & & 114 & 45 \\
1307 & Cimmeria & 63 & 9 & 2.820723 & 2 & 1 & 91 & 54 \\
1339 & Desagneauxa & 65 & 17 & 9.37510 & & & 78 & 120 \\
1396 & Outeniqua & 62 & 7 & 3.08175 & 2 & 1 & 112 & 68 \\
1493 & Sigrid & 78 & 7 & 43.179 &  &  & 78 & 103 \\
1619 & Ueta & 39 & 6 & 2.717943 & 5 & 1 & 122 & 51 \\
1623 & Vivian & $-$75 & 8 & 20.5235 & & & 77 & 58 \\
1738 & Oosterhoff & $-$72 & 8 & 4.44896 &  &  & 109 & 105 \\
1838 & Ursa & 47 & 17 & 16.1635 &  &  & 102 & 91 \\
2086 & Newell & $-$60 & 12 & 78.09 & 10 & 1 & 24 & 84 \\
3017 & Petrovic & $-$73 & 8 & 4.08037 & 3 & 1 & & 114 \\
3786 & Yamada & 56 & 2 & 4.03294 & 3 & 1 & & 71 \\
3896 & Pordenone & $-$32 & 9 & 4.00366 & 3 & 1 & 22 & 71 \\
4209 & Briggs & $-$56 & 25 & 12.2530 & 2 & 1 & & 64 \\
4467 & Kaidanovskij & 54 & 13 & 19.1454 & & & 20 & 107 \\
6179 & Brett & $-$42 & 20 & 9.4063 & 6 & 1 & & 93 \\
7055 & 1989 KB & $-$61 & 11 & 4.16878 & 7 & 1 & & 117 \\
7360 & Moberg & $-$18 & 18 & 4.58533 & 3 & 1 & & 103 \\
\hline
\end{longtable}
\tablefoot{For each asteroid, there is the mean ecliptic latitude $\beta$ of the pole direction and its dispersion $\Delta$, the other parameters have the same meaning as in Table \ref{tab:models}. The uncertainty of the sidereal rotational period corresponds to the last decimal place of $P$.
}

\begin{longtable}{r@{\,\,\,}l lcc}
\caption{\label{tab:references}Observations that are not included in the UAPC used for successful model determinations.}\\
\hline
 \multicolumn{2}{c} {Asteroid} & Date & Observer & Observatory (MPC code) \\ \hline\hline

\endfirsthead
\caption{continued.}\\

\hline
 \multicolumn{2}{c} {Asteroid} & Date & Observer & Observatory (MPC code) \\
\hline\hline
\endhead
\hline
\endfoot
364 & Isara & 2009 5 -- 2009 05 & \citet{Warner2009b} & Palmer Divide Observatory (716) \\
391 & Ingeborg & 2000 8 -- 2000 12 & \citet{Koff2001} & Antelope Hills Observatory, Bennett (H09) \\
502 & Sigune & 2007 6 -- 2007 6 & \citet{Stephens2007} & Goat Mountain Astronomical Research Station (G79) \\
553 & Kundry & 2004 12 -- 2005 1 & \citet{Stephens2005d} & Goat Mountain Astronomical Research Station (G79) \\
616 & Elly & 2010 1 -- 2010 1 & \citet{Warner2010a} & Palmer Divide Observatory (716) \\
 &  & 2010 2 -- 2010 2 & \citet{Durkee2010a} & Shed of Science Observatory, USA (H39) \\
621 & Werdandi & 2012 1 22.9 & \citet{Strabla2012a} & Bassano Bresciano Observatory (565) \\
 &  & 2012 1 -- 2012 2 & \citet{Strabla2012a} & Organ Mesa Observatory (G50) \\
1307 & Chimmeria & 2004 9 -- 2004 9 & \citet{Warner2005d} & Palmer Divide Observatory (716) \\
1396 & Outeniqua & 2006 3 -- 2006 3 & \citet{Warner2006a} & Palmer Divide Observatory (716) \\
1446 & Sillanpaa & 2009 3 -- 2009 3 & Higgins\tablefootmark{1} & Hunters Hill Observatory, Ngunnawal (E14) \\
1464 & Armisticia & 2008 1 -- 2008 1 & \citet{Brinsfield2008a} & Via Capote Sky Observatory, Thousand Oaks (G69) \\
1619 & Ueta & 2010 9 -- 2010 10 & \citet{Higgins2011a} & Hunters Hill Observatory, Ngunnawal (E14) \\
 &  & 2010 9 -- 2010 9 & \citet{Stephens2011a} & Goat Mountain Astronomical Research Station (G79) \\
1633 & Chimay & 2008 4 -- 2008 4 & \citet{Brinsfield2008b} & Via Capote Sky Observatory, Thousand Oaks (G69) \\
1987 & Kaplan & 2000 10 -- 2000 10 & \citet{Warner2001b,Warner2011a} & Palmer Divide Observatory (716) \\
 &  & 2011 12 -- 2011 12 & Warner & Palmer Divide Observatory (716) \\
2086 & Newell & 2007 1 -- 2007 2 & \citet{Stephens2007a} & Goat Mountain Astronomical Research Station (G79) \\
2403 & Bruce Helin & 2006 9 -- 2006 9 & Higgins\tablefootmark{1} & Hunters Hill Observatory, Ngunnawal (E14) \\
3279 & Solon & 2006 11 -- 2006 11 & \citet{Stephens2007b} & Goat Mountain Astronomical Research Station (G79) \\
3492 & Petra-Pepi & 2011 6 -- 2011 7 & \citet{Stephens2011b} & Goat Mountain Astronomical Research Station (G79) \\
3786 & Yamada & 2002 7 -- 2002 8 & \citet{Stephens2003b} & Goat Mountain Astronomical Research Station (G79) \\
3896 & Pordenone & 2007 10 -- 2007 10 & Higgins\tablefootmark{1} & Hunters Hill Observatory, Ngunnawal (E14) \\
4209 & Briggs & 2003 9 -- 2003 9 & \citet{Warner2004a} & Palmer Divide Observatory (716) \\
4399 & Ashizuri & 2008 6 -- 2008 6 & \citet{Brinsfield2008b} & Via Capote Sky Observatory, Thousand Oaks (G69) \\
4606 & Saheki & 2009 1 -- 2009 3 & \citet{Brinsfield2009a} & Via Capote Sky Observatory, Thousand Oaks (G69) \\
6159 & 1991 YH & 2006 3 -- 2006 3 & \citet{Warner2006a} & Palmer Divide Observatory (716) \\
6179 & Brett & 2009 4 -- 2009 4 & \citet{Warner2009a} & Palmer Divide Observatory (716) \\
6262 & Javid & 2010 2 -- 2010 2 & PTF\tablefootmark{2} &  \\
6403 & Steverin & 2004 9 -- 2004 9 & \citet{Warner2005d} & Palmer Divide Observatory (716) \\
7043 & Godart & 2008 8 -- 2008 8 & Durkee & Shed of Science Observatory, USA (H39) \\
 &  & 2008 8 -- 2008 9 & \citet{Pravec2012a} & Goat Mountain Astronomical Research Station (G79) \\
7055 & 1989 KB & 2007 5 -- 2007 5 & \citet{Stephens2007} & Goat Mountain Astronomical Research Station (G79) \\
 &  & 2007 5 -- 2007 6 & Higgins\tablefootmark{1} & Hunters Hill Observatory, Ngunnawal (E14) \\
7169 & Linda & 2006 8 -- 2006 8 & \citet{Higgins2007} & Hunters Hill Observatory, Ngunnawal (E14) \\
7360 & Moberg & 2006 4 -- 2006 4 & \citet{Oey2006a} & Leura (E17) \\
\end{longtable}
\tablefoot{
\tablefoottext{1}{On line at \texttt{http://www.david-higgins.com/Astronomy/asteroid/lightcurves.htm}} 
\tablefoottext{2}{Palomar Transient Factory survey \citep{Rau2009}, data taken from \citet{Polishook2012}.} 
}
\begin{longtable}{r@{\,\,\,}l cl}
\caption{\label{tab:interlopers}List of asteroids for which the HCM method alone suggest a membership to families Flora, Koronis, Eos, Eunomia, Phocaea and Alauda, but by additional methods for the family membership determination we identify them as interlopers or borderline cases.}\\
\hline 
\multicolumn{2}{c} {Asteroid} & \multicolumn{1}{c} {Status} & \multicolumn{1}{c} {Reason} \\
\hline\hline

\endfirsthead
\caption{continued.}\\

\hline
\multicolumn{2}{c} {Asteroid} & \multicolumn{1}{c} {Status} & \multicolumn{1}{c} {Reason} \\
\hline\hline
\endhead
\hline
\endfoot
\multicolumn{4}{c} {\textbf{Flora}} \\
9 & Metis & Interloper & Far from the ($a_{\mathrm{p}}$, $H$) border, peculiar SFD \\
43 & Ariadne & Interloper & Associated at $v_{\mathrm{cutoff}}=70$ m/s, peculiar SFD \\
352 & Gisela & Borderline & Associated at $v_{\mathrm{cutoff}}=70$ m/s, big object  \\
364 & Isara & Interloper & Big, peculiar SFD, close to ($a_{\mathrm{p}}$, $H$) border \\
376 & Geometria & Interloper & Far from the ($a_{\mathrm{p}}$, $H$) border, peculiar SFD \\
800 & Kressmannia & Borderline & Associated at $v_{\mathrm{cutoff}}=70$ m/s, lower albedo \\
1188 & Gothlandia & Borderline & Associated at $v_{\mathrm{cutoff}}=70$ m/s \\
1419 & Danzing & Interloper & Far from the ($a_{\mathrm{p}}$, $H$) border \\
1703 & Barry & Borderline & Associated at $v_{\mathrm{cutoff}}=70$ m/s \\
2839 & Annette & Interloper & Associated at $v_{\mathrm{cutoff}}=70$ m/s, C type \\
7360 & Moberg & Borderline & Redder (color from SDSS MOC4) \\ \hline
\multicolumn{4}{c} {\textbf{Koronis}} \\
167 & Urda & Borderline & Close to the ($a_{\mathrm{p}}$, $H$) border \\
208 & Lacrimosa & Interloper & Far from the ($a_{\mathrm{p}}$, $H$) border, peculiar SFD \\
311 & Claudia & Borderline & Close to the ($a_{\mathrm{p}}$, $H$) border \\
720 & Bohlinia & Borderline & Close to the ($a_{\mathrm{p}}$, $H$) border \\
1835 & Gajdariya & Interloper & Close to the ($a_{\mathrm{p}}$, $H$) border, incompatible albedo \\
2953 & Vysheslavia & Borderline & Close to the ($a_{\mathrm{p}}$, $H$) border \\
3170 & Dzhanibekov & Interloper & Behind the ($a_{\mathrm{p}}$, $H$) border, incompatible albedo \\ \hline
\multicolumn{4}{c} {\textbf{Eos}} \\
423 & Diotima & Interloper & Far from the ($a_{\mathrm{p}}$, $H$) border, big, C type \\
590 & Tomyris & Borderline & Close to the ($a_{\mathrm{p}}$, $H$) border \\ \hline
\multicolumn{4}{c} {\textbf{Eunomia}} \\
85 & Io & Interloper & Behind the ($a_{\mathrm{p}}$, $H$) border, peculiar SFD, incompatible albedo \\
390 & Alma & Borderline & Borderline albedo, borderline in ($a_{\mathrm{p}}$, $e_{\mathrm{p}}$, $I_{\mathrm{p}}$) space \\
4399 & Ashizuri & Borderline & Close to the ($a_{\mathrm{p}}$, $H$) border \\ \hline
\multicolumn{4}{c} {\textbf{Phocaea}} \\
290 & Bruna & Borderline & Close to the ($a_{\mathrm{p}}$, $H$) border \\
391 & Ingeborg & Interloper & Clearly outside ($a_{\mathrm{p}}$, $H$) \\
852 & Wladilena & Borderline & Slightly outside ($a_{\mathrm{p}}$, $H$) \\
1963 & Bezovec & Interloper & C type, incompatible albedo ($p_V$=0.04) \\
5647 & 1990 TZ & Interloper & Incompatible albedo ($p_V$=0.64) \\ \hline
\multicolumn{4}{c} {\textbf{Themis}} \\
62 & Erato & Borderline & Close to the ($a_{\mathrm{p}}$, $H$) border \\
1633 & Chimay & Borderline & Close to the ($a_{\mathrm{p}}$, $H$) border \\ \hline
\multicolumn{4}{c} {\textbf{Maria}} \\
695 & Bella & Borderline & Close to the ($a_{\mathrm{p}}$, $H$) border \\
714 & Ulula & Borderline & Close to the ($a_{\mathrm{p}}$, $H$) border \\ \hline
\multicolumn{4}{c} {\textbf{Alauda}} \\
276 & Adelheid & Interloper & Far from the ($a_{\mathrm{p}}$, $H$) border, big \\
\end{longtable}
\tablefoot{
In the table, we also give the name of the asteroid, the family membership according the HCM method, if it is an interloper or a borderline case and the reason. Peculiar SFD means a size frequency distribution that is incompatible with the SFD typically created by catastrophic collisions or cratering events (i.e., a large remnant, large fragment and steep slope). Quantity $v_{\mathrm{cutoff}}$ corresponds to the cutoff value of the HCM method for a particular family.
}

%
\scriptsize{
\begin{landscape}
\begin{longtable}{r@{\,\,\,}l rrrr D{.}{.}{6} rr ccccc}
\caption{\label{tab:families}List of asteroids that (i) have been identified as members of the Flora, Koronis, Eos, Eunomia, Phocaea, Themis, Maria, Vesta, Nysa/Polana and Alauda families  by the HCM method, and (ii) for which shape models from LI are available in the DAMIT database or are newly derived.}\\
\hline 
\multicolumn{2}{c} {Asteroid} & \multicolumn{1}{c} {$\lambda_1$} & \multicolumn{1}{c} {$\beta_1$} & \multicolumn{1}{c} {$\lambda_2$} & \multicolumn{1}{c} {$\beta_2$} & \multicolumn{1}{c} {$P$} & \multicolumn{1}{c} {$a_\mathrm{p}$} & \multicolumn{1}{c} {$D$} & \multicolumn{1}{c} {Bus/DeMeo} & \multicolumn{1}{c} {Tholen} & \multicolumn{1}{c} {$p_{\mathrm{V}}$} & \multicolumn{1}{c} {M/I/B} & \multicolumn{1}{c} {Reference} \\
\multicolumn{2}{l} { } & \multicolumn{1}{c} {[deg]} & \multicolumn{1}{c} {[deg]} & \multicolumn{1}{c} {[deg]} & \multicolumn{1}{c} {[deg]} & \multicolumn{1}{c} {[hours]} & \multicolumn{1}{c} {[AU]} & \multicolumn{1}{c} {[km]} &  &  &  &  &  \\
\hline\hline

\endfirsthead
\caption{continued.}\\

\hline
\multicolumn{2}{c} {Asteroid} & \multicolumn{1}{c} {$\lambda_1$} & \multicolumn{1}{c} {$\beta_1$} & \multicolumn{1}{c} {$\lambda_2$} & \multicolumn{1}{c} {$\beta_2$} & \multicolumn{1}{c} {$P$} & \multicolumn{1}{c} {$a_\mathrm{p}$} & \multicolumn{1}{c} {$D$} & \multicolumn{1}{c} {Bus/DeMeo} & \multicolumn{1}{c} {Tholen} & \multicolumn{1}{c} {$p_{\mathrm{V}}$} & \multicolumn{1}{c} {M/I/B} & \multicolumn{1}{c} {Reference} \\
\multicolumn{2}{l} { } & \multicolumn{1}{c} {[deg]} & \multicolumn{1}{c} {[deg]} & \multicolumn{1}{c} {[deg]} & \multicolumn{1}{c} {[deg]} & \multicolumn{1}{c} {[hours]} & \multicolumn{1}{c} {[AU]} & \multicolumn{1}{c} {[km]} &  &  &  &  &  \\
\hline\hline
\endhead
\hline
\endfoot
\multicolumn{14}{c} {\textbf{Flora}} \\
8 & Flora & 335 & $-$5 & 155 & 6 & 12.8667 & 2.2014 & 141.0 & $-$ & S & 0.26$\pm$0.05 & M & \citet{Torppa2003} \\
9 & Metis & 180 & 22 &  &  & 5.079177 & 2.3864 & 169.0 & $-$ & S & 0.13$\pm$0.02 & I & \citet{Torppa2003} \\
43 & Ariadne & 253 & $-$15 &  &  & 5.761987 & 2.2034 & 72.1 & Sk & S & 0.23$\pm$0.05 & I & \citet{Kaasalainen2002b} \\
281 & Lucretia & 128 & $-$49 & 309 & $-$61 & 4.349711 & 2.1878 & 11.8 & S & SU & 0.20$\pm$0.01 & M & \citet{Hanus2013a}/\citet{Kryszczynska2013a} \\
352 & Gisela & 205 & $-$26 & 23 & $-$20 & 7.48008 & 2.1941 & 26.7 & Sl & S & 0.19$\pm$0.02 & B & \citet{Hanus2013a} \\
364 & Isara & 282 & 44 & 86 & 42 & 9.15748 & 2.2208 & 35.2 & $-$ & S & 0.16$\pm$0.03 & I & this work \\
376 & Geometria & 239 & 45 & 63 & 53 & 7.71098 & 2.2886 & 39.0 & Sl & S & 0.19$\pm$0.04 & I & \citet{Hanus2011} \\
540 & Rosamunde & 301 & 81 & 127 & 62 & 9.34779 & 2.2189 & 20.3 & $-$ & S & 0.22$\pm$0.05 & M & this work \\
553 & Kundry & 197 & 73 & $-$1 & 64 & 12.6025 & 2.2308 & 9.6 & S & $-$ & 0.25$\pm$0.04 & M & this work \\
685 & Hermia & 197 & 87 & 29 & 79 & 50.387 & 2.2359 & 10.9 & $-$ & $-$ & 0.28$\pm$0.05 & M & \citet{Hanus2011} \\
700 & Auravictrix & 67 & 46 & 267 & 51 & 6.074836 & 2.2295 & 20.6 & $-$ & $-$ & 0.14$\pm$0.05 & M & \citet{Kryszczynska2013a} \\
800 & Kressmannia & 345 & 37 & 172 & 34 & 4.460964 & 2.1927 & 17.0 & $-$ & S & 0.15$\pm$0.02 & B & \citet{Hanus2011} \\
823 & Sisigambis & 86 & 74 &  &  & 146.58 & 2.2213 & 15.8 & $-$ & $-$ & 0.23$\pm$0.03 & M & \citet{Hanus2011} \\
915 & Cosette & 350 & 56 & 189 & 61 & 4.46974 & 2.2277 & 12.3 & $-$ & $-$ & 0.23$\pm$0.04 & M & \citet{Durech2009} \\
951 & Gaspra & 20 & 23 & 198 & 15 & 7.04203 & 2.2097 & 12.2 & S & S & 0.33$\pm$0.13 & M & this work \\
    &        & 19 & 21 &     &    &          &  &  &  &  &  &  & \citet{Davis1994}\footnotemark[1] \\
1056 & Azalea & 242 & 61 & 49 & 48 & 15.0276 & 2.2300 & 13.0 & S & $-$ & 0.25$\pm$0.04 & M & \citet{Hanus2013a} \\
1088 & Mitaka & 280 & $-$71 &  &  & 3.035378 & 2.2014 & 16.0 & S & S & 0.16$\pm$0.02 & M & \citet{Hanus2011} \\
1185 & Nikko & 359 & 34 &  &  & 3.786149 & 2.2375 & 11.3 & S & S & 0.20 & M & \citet{Hanus2011}/\citet{Durech2009} \\
1188 & Gothlandia & 133 & $-$84 & 335 & $-$81 & 3.491820 & 2.1907 & 12.7 & S & $-$ & 0.25$\pm$0.02 & B & \citet{Hanus2013a}/\citet{Kryszczynska2013a} \\
1249 & Rutherfordia & 204 & 72 & 31 & 74 & 18.2183 & 2.2243 & 14.1 & $-$ & S & 0.22$\pm$0.02 & M & \citet{Hanus2013a} \\
1270 & Datura & 60 & 76 &  &  & 3.358100 & 2.2347 & 8.2 & $-$ & $-$ & 0.24 & M & \citet{Vokrouhlicky2009} \\
1307 & Cimmeria &  & 63 &  &  & 2.820723 & 2.2505 & 10.1 & $-$ & S & 0.22$\pm$0.02 & B & this work \\
1396 & Outeniqua &  & 62 &  &  & 3.08175 & 2.2480 & 11.7 & $-$ & $-$ & 0.21$\pm$0.01 & M & this work \\
1419 & Danzig & 22 & 76 & 193 & 62 & 8.11957 & 2.2928 & 14.1 & $-$ & $-$ & 0.24$\pm$0.05 & I & \citet{Hanus2011} \\
1446 & Sillanpaa & 129 & 76 & 288 & 63 & 9.65855 & 2.2457 & 8.8 & $-$ & $-$ & 0.21$\pm$0.01 & M & this work \\
1514 & Ricouxa & 251 & 75 & 68 & 69 & 10.42467 & 2.2404 & 8.1 & $-$ & $-$ & 0.18$\pm$0.04 & M & \citet{Hanus2011} \\
1518 & Rovaniemi & 62 & 60 & 265 & 45 & 5.25047 & 2.2255 & 9.0 & $-$ & $-$ & 0.26$\pm$0.04 & M & \citet{Hanus2013a} \\
1527 & Malmquista & 274 & 80 &  &  & 14.0591 & 2.2274 & 10.3 & $-$ & $-$ & 0.22$\pm$0.02 & M & this work \\
1619 & Ueta &  & 39 &  &  & 2.717943 & 2.2411 & 9.9 & $-$ & S & 0.25$\pm$0.03 & M & this work \\
1675 & Simonida & 23 & 58 & 227 & 54 & 5.287962 & 2.2332 & 11.1 & $-$ & $-$ & 0.25$\pm$0.03 & M & \citet{Kryszczynska2013a} \\
1682 & Karel & 232 & 32 & 51 & 41 & 3.37486 & 2.2388 & 7.1 & $-$ & $-$ & 0.24 & M & \citet{Hanus2011} \\
1703 & Barry & 46 & $-$76 & 221 & $-$71 & 107.04 & 2.2148 & 9.4 & $-$ & $-$ & 0.22$\pm$0.03 & B & this work \\
1738 & Oosterhoff &  & $-$72 &  &  & 4.44896 & 2.1835 & 8.7 & S & $-$ & 0.28$\pm$0.04 & M & this work \\
1785 & Wurm & 11 & 57 & 192 & 47 & 3.26934 & 2.2359 & 6.2 & S & $-$ & 0.24 & M & \citet{Hanus2013a} \\
2017 & Wesson & 159 & 81 & 356 & 79 & 3.415579 & 2.2521 & 7.2 & $-$ & $-$ & 0.20$\pm$0.05 & M & \citet{Kryszczynska2013a} \\
2094 & Magnitka & 107 & 57 & 272 & 48 & 6.11219 & 2.2323 & 12.1 & $-$ & $-$ & 0.13$\pm$0.01 & M & \citet{Hanus2013a} \\
2112 & Ulyanov & 151 & 61 & 331 & 61 & 3.04071 & 2.2547 & 7.5 & $-$ & $-$ & 0.24 & M & \citet{Hanus2013a} \\
2510 & Shandong & 256 & 27 & 71 & 27 & 5.94638 & 2.2531 & 9.0 & $-$ & S & 0.20 & M & \citet{Hanus2013a} \\
2709 & Sagan & 308 & $-$8 & 124 & $-$16 & 5.25638 & 2.1954 & 6.8 & S & $-$ & 0.24 & M & \citet{Hanus2013a} \\
2839 & Annette & 341 & $-$49 & 154 & $-$36 & 10.4609 & 2.2166 & 7.6 & $-$ & $-$ & 0.06$\pm$0.01 & I & \citet{Hanus2013a} \\
3279 & Solon & 268 & $-$70 &  &  & 8.1041 & 2.2027 & 5.9 & $-$ & $-$ & 0.24 & M & this work \\
7043 & Godart & 73 & 62 & 235 & 80 & 8.4518 & 2.2447 & 5.7 & $-$ & $-$ & 0.23$\pm$0.04 & M & this work \\
7169 & Linda & 11 & $-$60 & 198 & $-$61 & 27.864 & 2.2487 & 4.5 & $-$ & $-$ & 0.24 & M & this work \\
7360 & Moberg &  & $-$18 &  &  & 4.58533 & 2.2510 & 7.7 & $-$ & $-$ & 0.22$\pm$0.04 & B & this work \\
31383 & 1998 XJ$_94$ & 110 & $-$74 & 279 & $-$63 & 4.16818 & 2.1853 & 4.1 & $-$ & $-$ & 0.29$\pm$0.03 & M & \citet{Hanus2013a} \\ \hline
\multicolumn{14}{c} {\textbf{Koronis}} \\
158 & Koronis & 30 & $-$64 &  &  & 14.2057 & 2.8687 &  47.7 &  S & S & 0.14$\pm$0.01 &  M &  \citet{Durech2011}  \\
    &      & 220 & $-$68 & 35 & $-$65 & 14.20569 &  &  &  &  &  &  & \citet{Slivan2003} \\
167 & Urda & 249 & $-$68 & 107 & $-$69 & 13.06133 & 2.8535 &  44.0 &  Sk & S & 0.16$\pm$0.04 &  B &  \citet{Durech2011}  \\
    &      & 225 & $-$73 & 40 & $-$75 & 13.06135 &  &  &  &  &  &  & \citet{Slivan2003} \\
208 & Lacrimosa & 170 & $-$68 & 350 & $-$71 & 14.076919 & 2.8929 &  45.0 &  Sk & S & 0.17$\pm$0.06 &  I &  \citet{Slivan2003}  \\
243 & Ida & 259 & $-$66 & 74 & $-$61 & 4.633632 & 2.8616 &  28.0 &  S & S & 0.24$\pm$0.07 &  M &  this work  \\
 &  & 263 & $-$67 &  &  & 4.633632 &  &  &  &  &  &  & \citet{Davies1994, Binzel1993}\footnotemark[2] \\
263 & Dresda & 105 & 76 & 285 & 80 & 16.81387 & 2.8865 &  25.5 &  S & $-$ & 0.18$\pm$0.02 &  M &  \citet{Slivan2009}  \\
277 & Elvira & 121 & $-$84 &  &  & 29.69219 & 2.8856 &  31.2 & $-$ & S & 0.20$\pm$0.05 &  M &  \citet{Hanus2011}  \\
    &      & 50 & $-$80 & 244 & $-$81 & 29.69218 &  &  &  &  &  &  & \citet{Slivan2009} \\
311 & Claudia & 214 & 43 & 30 & 40 & 7.5314 & 2.8976 &  25.8 & $-$ & S & 0.24$\pm$0.03 &  B &  \citet{Hanus2011}  \\
    &      & 209 & 48 & 24 & 48 & 7.53139 &  &  &  &  &  &  & \citet{Slivan2003} \\ 
321 & Florentina & 264 & $-$63 & 91 & $-$60 & 2.870866 & 2.8856 &  34.0 &  S & S & 0.14$\pm$0.01 &  M &  \citet{Slivan2003}  \\
462 & Eriphyla & 108 & 35 & 294 & 34 & 8.65890 & 2.8737 &  41.9 &  S & S & 0.17$\pm$0.02 &  M &  \citet{Slivan2009}  \\
534 & Nassovia & 66 & 41 & 252 & 42 & 9.46889 & 2.8842 &  38.6 &  Sq & S & 0.12$\pm$0.02 &  M &  \citet{Hanus2011}  \\
    &      & 58 & 50 & 244 & 50 & 9.46896 &  &  &  &  &  &  & \citet{Slivan2003} \\
720 & Bohlinia & 230 & 41 & 40 & 43 & 8.91862 & 2.8873 &  34.0 &  Sq & S & 0.20$\pm$0.02 &  B &  \citet{Slivan2003}  \\
832 & Karin & 242 & 46 & 59 & 44 & 18.35123 & 2.8644 &  16.3 &  $-$ & $-$ & 0.21$\pm$0.05 &  M &  \citet{Hanus2011}  \\
    &      & 230 & 42 & 52 & 42 & 18.352 &  &  &  &  &  &  & \citet{Slivan2012} \\
1223 & Neckar & 252 & 28 & 69 & 30 & 7.82401 & 2.8695 &  25.7 & $-$ & S & 0.15$\pm$0.03 &  M &  \citet{Hanus2011}  \\
    &      & 259 & 41 & 73 & 40 & 7.82124 &  &  &  &  &  &  & \citet{Slivan2003} \\
1289 & Kutaissi & 158 & $-$79 & 338 & $-$74 & 3.624174 & 2.8605 &  22.6 & $-$ & S & 0.16$\pm$0.04 &  M &  \citet{Slivan2003}  \\
1350 & Rosselia & 166 & $-$72 &  &  & 8.14011 & 2.8580 & 21.1 & Sa & S & 0.20$\pm$0.05 & M & \citet{Hanus2011} \\
1389 & Onnie & 183 & $-$75 & 360 & $-$79 & 23.0447 & 2.8661 & 14.7 & $-$ & $-$ & 0.17$\pm$0.04 & M & \citet{Hanus2013a} \\
1423 & Jose & 78 & $-$82 &  &  & 12.3127 & 2.8602 & 20.0 & S & $-$ & 0.28$\pm$0.04 & M & this work \\
1482 & Sebastiana & 262 & $-$68 & 91 & $-$67 & 10.48966 & 2.8723 & 17.6 & $-$ & $-$ & 0.21$\pm$0.05 & M & \citet{Hanus2011} \\
1618 & Dawn & 39 & $-$60 & 215 & $-$51 & 43.219 & 2.8688 & 17.5 & S & $-$ & 0.15$\pm$0.04 & M & this work \\
1635 & Bohrmann & 5 & $-$38 & 185 & $-$36 & 5.86427 & 2.8534 & 17.5 & S & $-$ & 0.21$\pm$0.02 & M & \citet{Hanus2011} \\
1742 & Schaifers & 56 & 52 & 247 & 68 & 8.53271 & 2.8892 & 16.6 & $-$ & $-$ & 0.11$\pm$0.02 & M & \citet{Hanus2011} \\
1835 & Gajdariya & 34 & 74 & 204 & 69 & 6.33768 & 2.8331 & 12.8 & $-$ & $-$ & 0.27$\pm$0.04 & I & this work \\
2953 & Vysheslavia & 11 & $-$64 & 192 & $-$68 & 6.29453 & 2.8282 & 12.8 & S & $-$ & 0.25$\pm$0.07 & B & \citet{Vokrouhlicky2006c} \\
3170 & Dzhanibekov & 216 & 62 & 30 & 63 & 6.07167 & 2.9291 & 9.6 & S & $-$ & 0.30$\pm$0.04 & I & \citet{Hanus2013a} \\
4507 & 1990 FV & 137 & 50 & 307 & 51 & 6.57932 & 2.8689 & 11.0 & $-$ & $-$ & 0.28$\pm$0.02 & M & \citet{Hanus2013a} \\
6262 & Javid & 93 & 76 & 275 & 66 & 8.02054 & 2.9063 & 7.8 & $-$ & $-$ & 0.29$\pm$0.04 & M & this work \\ \hline
\multicolumn{14}{c} {\textbf{Eos}} \\
423 & Diotima & 351 & 4 &  &  & 4.775377 & 3.0684 & 177.3 & C & C & 0.07$\pm$0.00 & I & \citet{Marchis2006} \\
573 & Recha & 74 & $-$24 & 252 & $-$48 & 7.16585 & 3.0138 & 44.4 & $-$ & $-$ & 0.13$\pm$0.02 & M & \citet{Hanus2011} \\
590 & Tomyris & 273 & $-$47 & 120 & $-$46 & 5.55247 & 3.0006 & 31.1 & $-$ & $-$ & 0.18$\pm$0.03 & B & \citet{Hanus2011} \\
669 & Kypria & 31 & 40 & 190 & 50 & 14.2789 & 3.0114 & 29.2 & $-$ & S & 0.17$\pm$0.02 & M & \citet{Hanus2013a} \\
807 & Ceraskia & 325 & 23 & 132 & 26 & 7.37390 & 3.0185 & 21.4 & $-$ & S & 0.21$\pm$0.05 & M & \citet{Hanus2013a} \\
1087 & Arabis & 334 & $-$7 & 155 & 12 & 5.79499 & 3.0150 & 45.6 & $-$ & S & 0.10$\pm$0.01 & M & \citet{Hanus2011} \\
1148 & Rarahu & 148 & $-$9 & 322 & $-$9 & 6.54448 & 3.0161 & 26.3 & K & S & 0.22$\pm$0.06 & M & \citet{Hanus2011} \\
1207 & Ostenia & 310 & $-$77 & 124 & $-$51 & 9.07129 & 3.0207 & 22.9 & $-$ & $-$ & 0.13$\pm$0.02 & M & \citet{Hanus2011} \\
1286 & Banachiewicza & 214 & 62 & 64 & 60 & 8.63041 & 3.0223 & 22.6 & $-$ & S & 0.16$\pm$0.03 & M & this work \\
1291 & Phryne & 106 & 35 & 277 & 59 & 5.58414 & 3.0130 & 22.4 & $-$ & $-$ & 0.19$\pm$0.04 & M & \citet{Hanus2011} \\
1339 & Desagneauxa &  & 65 &  &  & 9.37510 & 3.0211 & 26.1 & $-$ & S & 0.12$\pm$0.02 & M & this work \\
1353 & Maartje & 266 & 73 & 92 & 57 & 22.9927 & 3.0120 & 42.2 & $-$ & $-$ & 0.07$\pm$0.00 & M & this work \\
1464 & Armisticia & 194 & $-$54 & 35 & $-$69 & 7.46699 & 3.0035 & 23.3 & $-$ & $-$ & 0.13$\pm$0.36 & M & this work \\
2957 & Tatsuo & 88 & 57 & 246 & 37 & 6.82042 & 3.0221 & 22.9 & K & $-$ & 0.29$\pm$0.02 & M & \citet{Hanus2013a} \\
3896 & Pordenone &  & $-$32 &  &  & 4.00366 & 3.0057 & 20.0 & $-$ & $-$ & 0.13$\pm$0.01 & M & this work \\
5281 & Lindstrom & 238 & $-$72 & 84 & $-$81 & 9.2511 & 3.0125 & 20.0 & $-$ & $-$ &  & M & \citet{Hanus2013a} \\
19848 & Yeungchuchiu & 66 & $-$70 & 190 & $-$67 & 3.45103 & 3.0075 & 13.2 & $-$ & $-$ & 0.21$\pm$0.03 & M & \citet{Hanus2013a} \\ \hline
\multicolumn{14}{c} {\textbf{Eunomia}} \\
15 & Eunomia & 363 & $-$67 &  &  & 6.082752 & 2.6437 & 259.0 & S & S & 0.21$\pm$0.06 & M & \citet{Kaasalainen2002b} \\
85 & Io & 95 & $-$65 &  &  & 6.87478 & 2.6537 & 161.0 & B & FC & 0.06$\pm$0.03 & I & \citet{Durech2011} \\
390 & Alma & 54 & $-$48 & 263 & $-$73 & 3.74117 & 2.6517 & 31.2 & $-$ & DT & 0.13$\pm$0.02 & B & \citet{Hanus2013a} \\
812 & Adele & 301 & 44 & 154 & 69 & 5.85745 & 2.6594 & 13.6 & $-$ & $-$ & 0.24$\pm$0.03 & M & \citet{Hanus2013a} \\
1333 & Cevenola & 8 & $-$79 & 201 & $-$40 & 4.87933 & 2.6336 & 17.1 & $-$ & $-$ & 0.17$\pm$0.04 & M & \citet{Hanus2011} \\
1495 & Helsinki & 356 & $-$33 &  &  & 5.33131 & 2.6392 & 13.3 & $-$ & $-$ & 0.23$\pm$0.02 & M & \citet{Hanus2013a} \\
1503 & Kuopio & 170 & $-$86 & 27 & $-$61 & 9.9586 & 2.6263 & 18.4 & $-$ & $-$ & 0.30$\pm$0.06 & M & this work \\
1554 & Yugoslavia & 281 & $-$34 & 78 & $-$64 & 3.88766 & 2.6194 & 17.2 & $-$ & $-$ & 0.10$\pm$0.01 & M & \citet{Hanus2013a} \\
1927 & Suvanto & 90 & 39 & 277 & 6 & 8.16154 & 2.6497 & 12.5 & $-$ & $-$ & 0.26$\pm$0.04 & M & \citet{Hanus2013a} \\
2384 & Schulhof & 194 & $-$57 & 46 & $-$36 & 3.29367 & 2.6099 & 11.7 & $-$ & $-$ & 0.27$\pm$0.02 & M & \citet{Hanus2013a} \\
3017 & Petrovic &  & $-$73 &  &  & 4.08037 & 2.6074 & 12.7 & $-$ & $-$ & 0.21$\pm$0.02 & M & this work \\
3492 & Petra$-$Pepi & 9 & $-$57 & 202 & $-$16 & 46.570 & 2.6159 & 12.2 & $-$ & $-$ & 0.23$\pm$0.03 & M & this work \\
4399 & Ashizuri & 266 & $-$48 & 45 & $-$61 & 2.830302 & 2.5759 & 8.8 & $-$ & $-$ & 0.28$\pm$0.06 & B & this work \\
4467 & Kaidanovskij &  & 54 &  &  & 19.1454 & 2.6383 & 11.6 & $-$ & $-$ & 0.21 & M & this work \\
8132 & Vitginzburg & 33 & $-$66 & 193 & $-$48 & 7.27529 & 2.6263 & 11.6 & $-$ & $-$ & 0.21 & M & \citet{Hanus2013a} \\ \hline
\multicolumn{14}{c} {\textbf{Phocaea}} \\
25 & Phocaea & 347 & 10 &  &  & 9.935397 & 2.4002 & 75.1 & S & S & 0.23$\pm$0.02 & M & \citet{Hanus2013a} \\
290 & Bruna & 286 & $-$80 & 37 & $-$74 & 13.8055 & 2.3372 & 10.4 & $-$ & $-$ & 0.42$\pm$0.08 & B & \citet{Hanus2013a} \\
391 & Ingeborg &  & $-$60 &  &  & 26.4145 & 2.3202 & 19.6 & S & S & 0.20 & I & this work \\
502 & Sigune &  & $-$44 &  &  & 10.92667 & 2.3831 & 19.5 & $-$ & S & 0.23$\pm$0.02 & M & this work \\
852 & Wladilena & 218 & $-$41 & 57 & $-$16 & 4.613301 & 2.3627 & 31.1 & $-$ & $-$ & 0.16$\pm$0.02 & B & \citet{Hanus2013a} \\
1192 & Prisma &  & $-$65 &  &  & 6.55836 & 2.3660 & 7.2 & $-$ & $-$ & 0.23 & M & this work \\
1568 & Aisleen & 109 & $-$68 &  &  & 6.67598 & 2.3520 & 12.0 & $-$ & $-$ & 0.18$\pm$0.03 & M & \citet{Hanus2011} \\
1963 & Bezovec & 218 & 16 & 50 & $-$49 & 18.1655 & 2.4231 & 45.0 & $-$ & C & 0.04$\pm$0.01 & I & \citet{Hanus2013a} \\
1987 & Kaplan & 356 & $-$58 &  &  & 9.45950 & 2.3822 & 14.6 & $-$ & $-$ & 0.21$\pm$0.04 & M & this work \\
2430 & Bruce Helin & 177 & $-$68 &  &  & 129.75 & 2.3627 & 12.7 & Sl & S & 0.23 & M & this work \\
5647 & 1990 TZ & 266 & 69 &  &  & 6.13868 & 2.4241 & 9.3 & S & $-$ & 0.64$\pm$0.07 & I & \citet{Hanus2013a} \\
6179 & Brett &  & $-$42 &  &  & 9.4063 & 2.4278 & 5.8 & $-$ & $-$ & 0.23 & M & this work \\
7055 & 1989 KB &  & $-$61 &  &  & 4.16878 & 2.3496 & 6.7 & $-$ & $-$ & 0.33$\pm$0.15 & M & this work \\
10772 & 1990 YM & 16 & 46 &  &  & 68.82 & 2.3901 & 6.2 & $-$ & $-$ & 0.38$\pm$0.06 & M & \citet{Hanus2013a} \\ \hline
\multicolumn{14}{c} {\textbf{Themis}} \\
62 & Erato & 87 & 22 & 269 & 23 & 9.21813 & 3.1217 & 95.4 & Ch & BU & 0.06$\pm$0.00 & B & \citet{Hanus2011} \\
222 & Lucia & 107 & 54 & 290 & 51 & 7.83671 & 3.1349 & 56.5 & $-$ & BU & 0.12$\pm$0.02 & M & \citet{Hanus2013a} \\
621 & Werdandi & 247 & $-$86 & 66 & $-$77 & 11.77456 & 3.1193 & 27.1 & $-$ & FCX & 0.15$\pm$0.02 & M & this work \\
936 & Kunigunde & 47 & 57 & 234 & 50 & 8.82653 & 3.1383 & 39.6 & $-$ & $-$ & 0.11$\pm$0.01 & M & this work \\
1003 & Lilofee &  & 65 &  &  & 8.24991 & 3.1483 & 31.4 & $-$ & $-$ & 0.15$\pm$0.04 & M & this work \\
1623 & Vivian &  & $-$75 &  &  & 20.5235 & 3.1347 & 29.6 & $-$ & $-$ & 0.08 & M & this work \\
1633 & Chimay & 322 & 77 & 116 & 81 & 6.59064 & 3.1748 & 37.7 & $-$ & $-$ & 0.08$\pm$0.01 & B & this work \\
1691 & Oort & 45 & 68 & 223 & 58 & 10.2684 & 3.1664 & 33.2 & $-$ & CU & 0.07$\pm$0.01 & M & this work \\
1805 & Dirikis & 364 & 48 & 188 & 61 & 23.4543 & 3.1333 & 28.1 & $-$ & $-$ & 0.09$\pm$0.01 & M & this work \\ \hline
\multicolumn{14}{c} {\textbf{Maria}} \\
616 & Elly &  & 67 &  &  & 5.29771 & 2.5526 & 22.6 & $-$ & S & 0.19$\pm$0.04 & M & this work \\
695 & Bella & 87 & $-$55 & 314 & $-$56 & 14.21899 & 2.5391 & 41.2 & $-$ & S & 0.24$\pm$0.03 & B & \citet{Hanus2011} \\
714 & Ulula & 224 & $-$10 & 41 & $-$5 & 6.99838 & 2.5352 & 39.2 & $-$ & S & 0.27$\pm$0.04 & B & \citet{Hanus2011} \\
787 & Moskva & 330 & 60 & 122 & 19 & 6.05581 & 2.5396 & 40.3 & $-$ & $-$ & 0.12$\pm$0.02 & M & \citet{Hanus2013a} \\
875 & Nymphe & 42 & 31 & 196 & 42 & 12.6213 & 2.5539 & 15.2 & $-$ & $-$ & 0.19$\pm$0.02 & M & \citet{Hanus2013a} \\
1160 & Illyria &  & 47 &  &  & 4.10295 & 2.5604 & 14.8 & $-$ & $-$ & 0.22$\pm$0.04 & M & this work \\
1996 & Adams & 107 & 55 &  &  & 3.31114 & 2.5587 & 13.5 & $-$ & $-$ & 0.14$\pm$0.01 & M & \citet{Hanus2013a} \\
3786 & Yamada &  & 56 &  &  & 4.03294 & 2.5503 & 16.7 & $-$ & $-$ & 0.23$\pm$0.04 & M & this work \\
6403 & Steverin & 246 & 77 & 109 & 73 & 3.49119 & 2.5945 & 6.9 & $-$ & $-$ & 0.49$\pm$0.05 & M & this work \\ \hline
\multicolumn{14}{c} {\textbf{Vesta}} \\
63 & Ausonia & 305 & $-$21 & 120 & $-$15 & 9.29759 & 2.3952 & 90.0 & Sa & S & 0.16$\pm$0.03 & $-$ & \citet{Torppa2003} \\
306 & Unitas & 79 & $-$35 &  &  & 8.73874 & 2.3580 & 49.0 & S & S & 0.17$\pm$0.06 & $-$ & \citet{Durech2007} \\
336 & Lacadiera & 194 & 39 & 37 & 54 & 13.69555 & 2.2518 & 69.0 & Xk & D & 0.05$\pm$0.01 & $-$ & \citet{Hanus2011} \\
556 & Phyllis & 34 & 54 & 209 & 41 & 4.292622 & 2.4654 & 38.5 & S & S & 0.18$\pm$0.03 & $-$ & \citet{Marciniak2007} \\
1933 & Tinchen & 113 & 26 & 309 & 36 & 3.67062 & 2.3530 & 6.5 & $-$ & $-$ & 0.29$\pm$0.06 & $-$ & \citet{Hanus2013a} \\
2086 & Newell &  & $-$60 &  &  & 78.09 & 2.4014 & 9.8 & Xc & $-$ & 0.20 & $-$ & this work \\
6159 & 1991 YH & 266 & 67 & 62 & 67 & 10.6589 & 2.2914 & 5.4 & $-$ & $-$ & 0.46$\pm$0.13 & $-$ & this work \\
8359 & 1989 WD & 121 & $-$68 & 274 & $-$68 & 2.89103 & 2.3500 & 8.2 & $-$ & $-$ & 0.22$\pm$0.03 & $-$ & \citet{Hanus2013a} \\ \hline
\multicolumn{14}{c} {\textbf{Nysa/Polana}} \\
44 & Nysa & 99 & 58 &  &  & 6.421417 & 2.4227 & 70.6 & Xc & E & 0.55$\pm$0.07 & $-$ & \citet{Kaasalainen2002b} \\
135 & Hertha & 272 & 52 &  &  & 8.40060 & 2.4285 & 77.0 & Xk & M & 0.15$\pm$0.05 & $-$ & \citet{Torppa2003} \\
1378 & Leonce & 210 & $-$67 & 46 & $-$77 & 4.32526 & 2.3748 & 22.5 & $-$ & $-$ & 0.03$\pm$0.00 & $-$ & this work \\
1493 & Sigrid &  & 78 &  &  & 43.179 & 2.4297 & 22.1 & Xc & F & 0.04$\pm$0.00 & $-$ & this work \\
4606 & Saheki & 44 & 59 & 222 & 68 & 4.97347 & 2.2518 & 6.7 & $-$ & $-$ & 0.33$\pm$0.02 & $-$ & this work \\ \hline
\multicolumn{14}{c} {\textbf{Alauda}} \\
276 & Adelheid & 199 & $-$20 & 9 & $-$4 & 6.319200 & 3.1162 & 125.0 & $-$ & X & 0.06$\pm$0.01 & I & \citet{Marciniak2007} \\
1276 & Ucclia &  & $-$49 &  &  & 4.90748 & 3.1698 & 40.0 & $-$ & $-$ & 0.05$\pm$0.01 & M & this work \\
1838 & Ursa &  & 47 &  &  & 16.1635 & 3.2111 & 48.6 & $-$ & $-$ & 0.04$\pm$0.01 & M & this work \\
4209 & Briggs &  & $-$56 &  &  & 12.2530 & 3.1564 & 30.9 & $-$ & $-$ & 0.09$\pm$0.03 & M & this work \\
\hline
\end{longtable}
\tablefoot{
For each asteroid, the table gives the spin state solution (i.e., ecliptic coordinates $\lambda$ and $\beta$ of the spin axis and the sidereal rotational period $P$, usually for both ambiguous pole solutions), the proper semi-major axis $a_\mathrm{p}$, the diameter $D$ and albedo $p_{\mathrm{v}}$ based on WISE data \citep{Masiero2011}, the SMASS II taxonomy \citep{Bus2002}, the Tholen \citep{Tholen1984,Tholen1989} taxonomical type, the information if the asteroid is, according to our membership revision, a member (M), an interloper (I) or a borderline case (B), and the reference to the convex model.\\
\tablefoottext{1}{The spin vector solution of asteroid (951) Gaspra is based on Galileo images obtained during the October 1991 flyby.}
\tablefoottext{2}{The solution of asteroid (243) Ida is based on Galileo images and photometric data.} }
\end{landscape}
}

\end{document}